\documentclass[aps,prb,notitlepage,twocolumn,amsmath,amssymb,10pt]{revtex4-1}
\usepackage{times}

\usepackage[usenames,dvipsnames]{xcolor}

\usepackage{tikz}  
\usetikzlibrary{arrows,shapes,positioning,shadows,backgrounds,fit}

\usepackage[english]{babel}
\usepackage{graphicx, dcolumn, bm, color, amsmath, relsize, amsmath, dsfont, mathrsfs, empheq, verbatim, upgreek, etoolbox,braket}
\usepackage[caption = false]{subfig}

\usepackage[colorlinks=true,linkcolor=blue,urlcolor=blue,citecolor=blue]{hyperref}
\usepackage{graphics}
\usepackage{bm}
\usepackage{color}
\usepackage{amscd}
\usepackage{amsfonts}
\usepackage{amssymb}
\usepackage{graphicx}
\usepackage{tabularx}

\newcommand{\mean}[1]{\langle #1 \rangle}
\providecommand{\bgreek}[1]{\mbox{\boldmath$#1$}}

\newcommand{\Ham}{\mathcal{H}}
\newcommand{\Id}{\mathds{1}}

\newcommand{\e}{\eta}
\newcommand{\ii}{{\rm i}}

\allowdisplaybreaks

\newcommand{\GGrev}[1]{\textcolor{black}{#1}}
\newcommand{\SCrev}[1]{\textcolor{black}{#1}}

\definecolor{myblue}{rgb}{.8, .8, 1}

\newlength\mytemplen
\newsavebox\mytempbox

\makeatletter
\newcommand\mybluebox{%
    \@ifnextchar[
       {\@mybluebox}%
       {\@mybluebox[0pt]}}

\def\@mybluebox[#1]{%
    \@ifnextchar[
       {\@@mybluebox[#1]}%
       {\@@mybluebox[#1][0pt]}}

\def\@@mybluebox[#1][#2]#3{
    \sbox\mytempbox{#3}%
    \mytemplen\ht\mytempbox
    \advance\mytemplen #1\relax
    \ht\mytempbox\mytemplen
    \mytemplen\dp\mytempbox
    \advance\mytemplen #2\relax
    \dp\mytempbox\mytemplen
    \colorbox{myblue}{\hspace{1em}\usebox{\mytempbox}\hspace{1em}}}

\begin{document}

\newdimen\origiwspc%
\newdimen\origiwstr%
\preprint{}

\title{Thermodynamics of precision in quantum non-equilibrium steady states}
\author{Giacomo Guarnieri$^{1}$}
\email{guarnieg@tcd.ie}
\author{Gabriel T. Landi$^2$}
\email{gtlandi@if.usp.br}
\author{Stephen R. Clark$^{3,4}$}
\email{stephen.clark@bristol.ac.uk}
\author{John Goold$^{1}$}
\email{gooldj@tcd.ie}
\affiliation{$^1$\mbox{Department of Physics, Trinity College Dublin, Dublin 2, Ireland}}
\affiliation{$^2$\mbox{Instituto de F\'isica da Universidade de S\~ao Paulo,  05314-970 S\~ao Paulo, Brazil}}
\affiliation{$^3$\mbox{H.H. Wills Physics Laboratory, University of Bristol, Bristol BS8 1TL, UK.}}
\affiliation{$^4$\mbox{Max Planck Institute for the Structure and Dynamics of Matter,
University of Hamburg CFEL, Hamburg, Germany.}}

\begin{abstract}

Autonomous engines operating at the nano-scale can be prone to deleterious fluctuations in the heat and particle currents which increase, for fixed power output, the more reversible the operation regime is. 
This fundamental trade-off between current fluctuations and entropy production forms the basis of the so-called thermodynamic uncertainty relations (TURs).
\GGrev{Importantly, recent studies have shown that they can be violated in the quantum regime, thus motivating the search for analogous quantum counterparts. 
In this paper we show that the geometry of quantum non-equilibrium steady-states alone directly implies the existence of TUR, but with a looser bound, which is not violated by the above recent findings.}
The geometrical nature of this result makes it extremely general, establishing a fundamental limit for the thermodynamics of precision.
Our proof is based on the McLennan-Zubarev ensemble, which provides an exact description of non-equilibrium steady-states.  
We first prove that  the entropy production of this ensemble can be expressed as a quantum relative entropy.
The TURs are then shown to be a direct consequence of the Cramer-Rao bound, a fundamental result from parameter estimation theory.
By combining techniques from many-body physics and information sciences, our approach also helps to shed light on the delicate relationship between quantum effects and current fluctuations in autonomous machines, where new general bound on the power output are found and discussed. 
\end{abstract}
\date{\today}
\maketitle

\section{Introduction}
\label{sec:intro}

Autonomous machines, whether classical or quantum, generically operate in  non-equilibrium conditions. 
They harvest work by consuming resources such as heat or fuel and, in order to maintain functionality, dissipate into the environment~\cite{benenti2017fundamental}. 
As such they operate in a regime known as non-equilibrium steady state (NESS), characterized by a non-zero entropy production rate~\cite{de2013non,VonOppenPRL} and by an ability to maintain non-zero average currents across the system. Accurately describing the physical properties of a NESS is central to the development of mesoscopics ~\cite{imry2002introduction,akkermans2007mesoscopic} as well as fundamental to our understanding of nano-scale autonomous machines such as molecular electronics~\cite{nitzan2003electron,galperin2007molecular}, nano-junction thermoelectrics~\cite{dubi2011colloquium}, single electron circuits \cite{pekola2018thermodynamics}, quantum dots~\cite{josefsson2018quantum}, quantum autonomous refrigerators~\cite{refrigerator1,refrigerator2,refrigerator3,mitchison2019quantum} and even ultra-cold atomic systems~\cite{krinner2017two}.

An important advance in this respect has been made recently  with the discovery that classical time-homogeneous Markovian chains obey the so-called thermodynamic uncertainty relations (TURs)~\cite{barato2015thermodynamic,gingrich2016dissipation}.
The basic idea is that in meso- and microscopic systems, fluctuations of the currents around their mean values become significant. 
The TUR provides a bound on these fluctuations by relating them to the NESS entropy production rate according to (taking $k_B = 1$ and $\hbar = 1$ throughout)
\begin{equation}\label{classicalTUR}
\frac{\Delta_{\hat{J}_\alpha}}{\langle \hat{J}_\alpha \rangle^2} \langle \hat{\sigma}\rangle\geq 2,
\end{equation}
where $\langle \hat{J}_\alpha\rangle$ is the average current (of particles, charge or heat), $\Delta_{\hat{J}_\alpha} \equiv \lim_{\mathfrak{T}\to \infty} \mathfrak{T}\left(\langle \hat{J}_\alpha^2\rangle-\langle \hat{J}_\alpha\rangle^2\right)$ denotes its normalized variance and $\langle \hat{\sigma} \rangle$ is the average entropy production rate in the NESS.
\\Since the original inception of this result, there has been a flurry of activity aimed at further exploring their consequences in various settings~\cite{polettini2016tightening,nardini2018process,horowitz2017proof,
pietzonka2017finite,Macieszczak2018PRL,
proesmans2019hysteretic,potts2019thermodynamic,Timpa2019,
liu2019thermodynamic,koyuk2019operationally,miller2019work,
hasegawa2019generalized,van2019uncertainty}. Not only is the TUR expected to have implications for the functioning of biological clocks~\cite{barato2016cost} and control techniques~\cite{barato2017thermodynamic}, it was demonstrated recently that it has significant consequences for the operation of autonomous machines. 
For instance, it follows from Eq.~\eqref{classicalTUR} that the fluctuations in the output power of an engine operating between two reservoirs at temperatures $T_C$ and $T_H > T_C$ is bounded by  \cite{Pietzonka2018PRL}
\begin{equation}\label{eq:BPS}
\mean{\hat{P}} \leq \frac{\Delta_P}{2T_C}\left(\frac{\eta_C}{\eta} - 1\right) \equiv \mathcal{B}_{PS},
\end{equation}
where $\mean{\hat{P}}$ denotes the average power, $\Delta_P$ its normalized fluctuations, $\eta$ the engine's efficiency and $\eta_C = 1- T_C/T_H$ the corresponding Carnot efficiency. 
This result implies that, in order to have an autonomous steady-state machine which operates at finite power as $\eta \to \eta_C$, one must incur fluctuations that diverge at least as $\sim (\eta_C-\eta)^{-1}$.

Recently, it has been shown \cite{Agarwalla2018PRB, Ptaszynski2018PRB,liu2019thermodynamic} that the classical TUR Eq.~(\ref{classicalTUR}) can  be violated in the quantum regime. 
While the precise mechanisms responsible for these violations are still not fully understood, this opens up an interesting perspective, as it would in principle allow one to use quantum effects to reduce the deleterious current fluctuations without compromising the engine's efficiency and output power \cite{kosloff2013quantum,goold2016role,vinjanampathy2016quantum}. 
Moreover, these violations also naturally lead one to ask, to what extent, are TURs really a universal feature of non-equilibrium steady-states. 

In this work we show that the geometry of quantum NESS, by itself, already implies the existence of a TUR of the form
\begin{equation}\label{quantumTUR}
\mean{\hat{\sigma}} \geq \mean{\hat{\mathbf{J}}}^T \bgreek{\Delta}^{-1} \mean{\hat{\mathbf{J}}},
\end{equation}
where $\bgreek{\Delta} $ is the normalized covariance matrix between different steady-state currents. By restricting to the single-component vector case, one immediately in particular obtains
\begin{equation}
\frac{\Delta_{\hat{J}_\alpha}}{\mean{\hat{J}_\alpha}^2}\mean{\hat{\sigma}} \geq 1,
\end{equation}
which, compared to the classical result in Eq.~(\ref{classicalTUR}), shows that our bound involving the variance of currents can in principle be \GGrev{two} times looser.  
The key to achieve this result is, rather than adopting a dynamical approach, to exploit the generalization of the idea of Gibbs distributions to the set of NESSs known as the non-equilibrium statistical operator approach, or the McLennan-Zubarev form ~\cite{Zubarev1,McLennan1,Hershfield}. This description allows us to write the average entropy production in the NESS as a quantum relative entropy and investigate the geometrical structure of the manifold defined by this family of states. Making use of concepts borrowed from the geometry of quantum states and quantum estimation theory, this allows us to derive an expression of the steady-state entropy production in terms of a relative entropy with a positive correction depending on powers of the current fluctuations in equilibrium, and then the geometric bound Eq.~\eqref{quantumTUR}. We finally illustrate the implications of our findings to the output power of an autonomous mesocopic heat engines, in the same spirit as Ref.~\cite{barato2017thermodynamic}. This leads to the following new general independent upper bounds
\begin{align}\label{powerbounds}
    &\mean{\hat{P}} \leq 2 \mathcal{B}_{PS}, \notag\\
    &\mean{\hat{P}} \leq \frac{\eta}{T_C} \frac{\Delta_{\hat{P}}\Delta_{\hat{J}_H} - \Delta_{\hat{P}, \hat{J}_H}}{\Delta_{\hat{P}} - 2\eta\Delta_{\hat{P}, \hat{J}_H} + \eta^2 \Delta_{\hat{J}_H}} \left(\eta_c - \eta\right),
\end{align}
where $\mathcal{B}_{PS}$ is given by Eq.~\eqref{eq:BPS}, $\Delta_{\hat{P}, \hat{J}_H}$ denotes the normalized correlation between the power and the heat current from the hot reservoir, while $\Delta_{\hat{J}_H}$ is the variance of the latter.
On the one hand, one can deduce from the first inequality in Eq.~\eqref{powerbounds} that the maximum reachable power for a nanoscale steady-state engine \GGrev{can be two} times larger than any classical Markovian counterpart.
Furthermore, our generalized quantum TUR \SCrev{both reveals and quantifies} how much the fluctuations of the incoming heat current from the hot reservoir affects the achievable output power. 
Finally, we show a concrete application of our results in a paradigmatic toy model consisting of a serial double quantum dot connected to two \SCrev{semi-infinite} fermionic leads. A more systematic study of these two bounds in other physical platforms and models will be pursued in a forthcoming work.

\section{Non-equilibrium steady-state statistical operator}
\label{sec:ness}

\begin{figure}[htbp!]
\begin{center}
\begin{tikzpicture} 
  \node (img1)  {\includegraphics[scale=0.4]{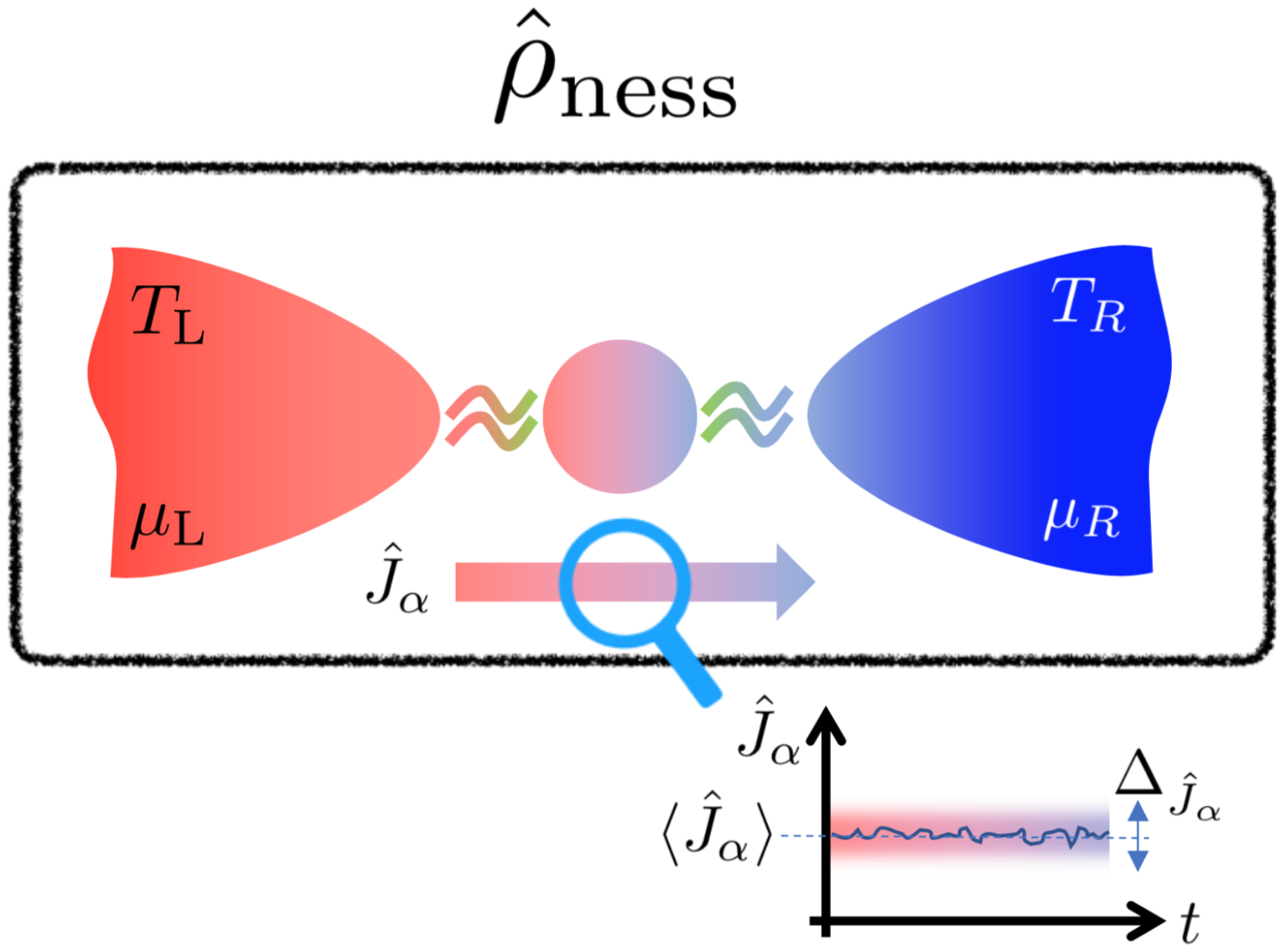}};
    \node[above=of img1, node distance=0cm, yshift=-1.7cm,xshift=-2.1cm] {{\color{black}{\bf{(a)}}}};
\end{tikzpicture}\\
\begin{tikzpicture} 
  \node (img1)  {\includegraphics[scale=0.4]{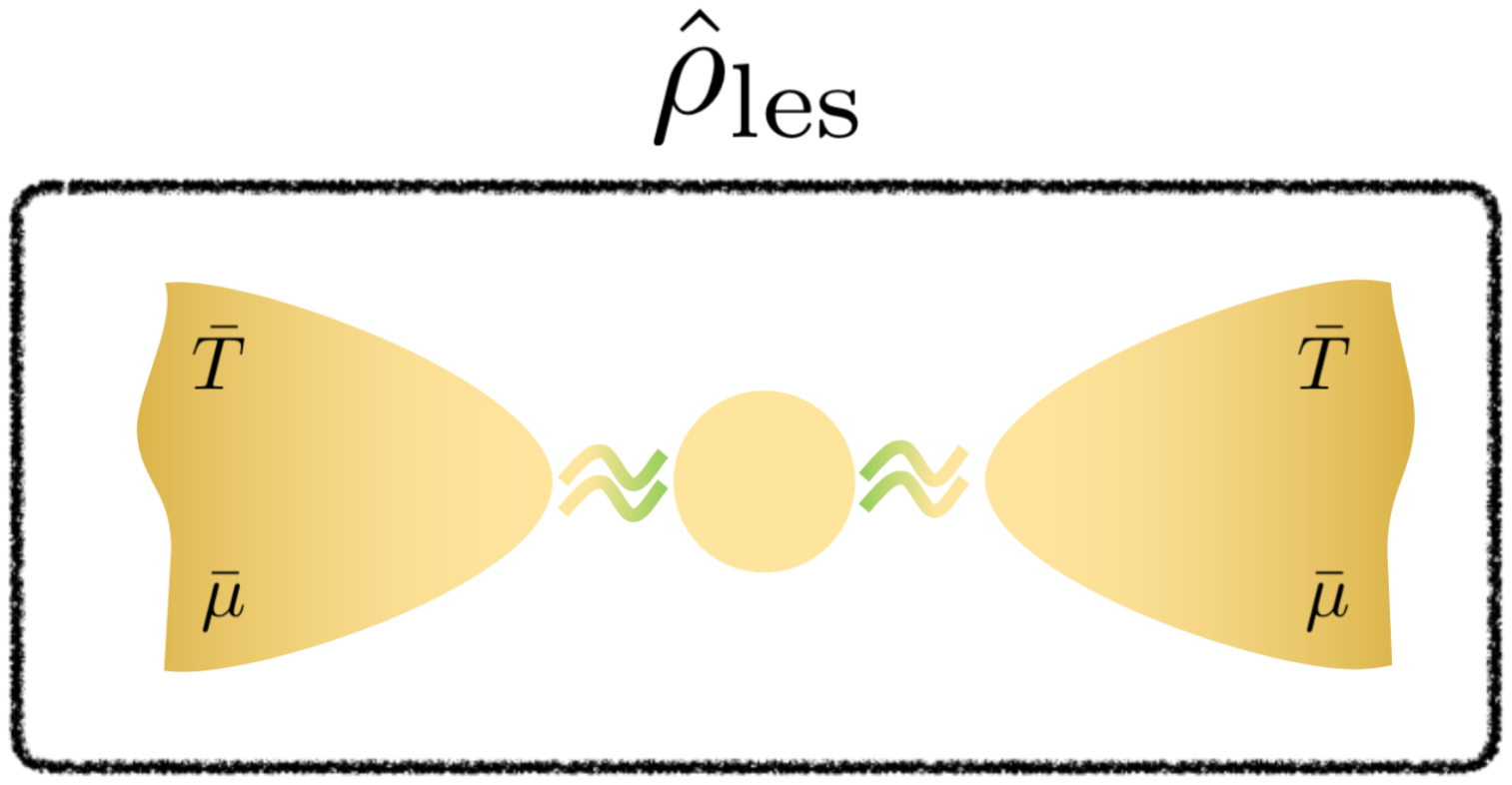}};
    \node[above=of img1, node distance=0cm, yshift=-1.7cm,xshift=-2.1cm] {{\color{black}{\bf{(b)}}}};
\end{tikzpicture}
\vspace*{-0.2cm}
\caption{Diagrammatic illustration of the scenario considered in this paper. 
A central system ($C$) is coupled to two semi-infinite thermal reservoirs prepared at temperatures $T_{L,R}$ and chemical potentials $\mu_{L,R}$. 
After a sufficiently long time, this system will tend to a global non-equilibrium steady-state (NESS) {\bf{(a)}} characterized by the existence of a finite current of particles ($\hat{J}_Q$) and energy ($\hat{J}_E$) through the central system. When the two biases in temperature and chemical potential are set to zero, i.e $T_L = T_R =\overline{T}$ and $\mu_L = \mu_R = \overline{\mu}$, the asymptotic state becomes an equilibrium state, denoted as a local equilibrium state (LES) {\bf{(b)}}. The two states are closely related to each other as shown in Eq.~\eqref{eq:NessLes}.} 
\label{figStates}
\end{center}
\end{figure}

\SCrev{For clarity, we will consider a typical NESS scenario depicted in Fig.~\ref{figStates}(a), whereby a central quantum system is connected to two semi-infinite fermionic leads, $L$ and $R$, acting both as energy and particle reservoirs that  drive the overall system into a global steady-state. The results we derive, however, can be straightforwardly generalized to the case of multiple bosonic and/or fermionic baths.} This extension is discussed in more detail in Appendix~\ref{SubsecSI:DerivNess1} and in Ref.~\cite{ZubarevBook1}, while a derivation from a purely classical perspective is given in Ref.~\cite{McLennan1}.

The main idea of the statistical operator approach is to use a density matrix ensemble description for the NESS, taking into account the additional conserved quantities which are responsible for the currents. This is based on propagating the entire composite system from the infinite past to the present using a generalized Gibbs ensemble first derived by McLennan \cite{McLennan1, McLennan2} and Zubarev \cite{Zubarev1, Zubarev2, Zubarev3}, and is hence known as the McLennan-Zubarev form. An alternative form for the NESS statistical operator was derived in the 90's by Hershfield \cite{Hershfield} using a scattering-theory based approach. Not only are the approaches known to be equivalent~\cite{Ness1,Ness2,Ness3} but they can also can be obtained from a max-entropy approach~\cite{Bokes1, Bokes2}, \SCrev{analogous to equilibrium ensembles~\cite{Jaynes1,Jaynes2}, but with constrained finite currents.}

\SCrev{For the NESS scenario outlined the overall system} is described by the total Hamiltonian $\hat{\Ham} = \hat{\Ham}_0 + \hat{\Ham}_{\mathrm{int}}$, with $\hat{\Ham}_0 = \hat{\Ham}_C + \hat{\Ham}_L + \hat{\Ham}_R$ representing the sum of all the autonomous terms for each part, and $\hat{\Ham}_{\mathrm{int}} = \hat{V}_{\mathrm{LC}} + \hat{V}_{\mathrm{RC}} $ incorporating all the couplings between the three parts. \SCrev{We make no assumptions about the contributions comprising $\hat{\Ham}$ besides that they conserve the particle number. In particular the central conductor $C$ could itself be a complex many-body system.} 

At the infinite past, $t_0 = -\infty$,  the three components are considered decoupled, with the two baths being at equilibrium characterized by two different inverse temperatures $\beta_L$ and $\beta_R$ and two chemical potentials $\mu_L$ and $\mu_R$. The total system is therefore taken to be at the state $\hat{\rho}(t_0) =\hat{ \rho}_L \otimes \hat{\rho}_C \otimes \hat{\rho}_R,$ where
\begin{equation}\label{rho0}
   \hat{\rho}_a = Z_{a}^{-1} e^{-\beta_{a} \left(\hat{\Ham}_a - \mu_a \hat{N}_a\right)}, \qquad \qquad (a = L,R),
\end{equation}
in which \GGrev{$Z_{a} = \mathrm{Tr}_{a}[ e^{-\beta_{a} \left(\hat{\Ham}_a - \mu_a \hat{N}_a\right)}] $} and  $\hat{N}_a$ is the total particle number operator for reservoir $a$. We are interested in the steady-state and so  any observable of interest in this limit will be independent from the initial state of the central system $\hat{\rho}_C$.

Immediately after the initial time $t_0$ the coupling between the central system and the two leads is switched-on adiabatically according to $\hat{\Ham}_{\epsilon}(t) = \hat{\Ham}_0 + e^{-\epsilon |t|} \hat{\Ham}_{\mathrm{int}}$, with $\epsilon$ being an arbitrary small positive constant~\cite{FetterWalecka}, so $\hat{\Ham}_{\epsilon} (t_0 = -\infty) = \hat{\Ham}_0$, $\hat{\Ham}_{\epsilon} (0) = \hat{\Ham}$. Making use of the interaction picture evolution operator $\hat{U}_{I,\epsilon}(0, -\infty)$ generated by this Hamiltonian, the NESS statistical operator $\hat{\rho}_{\mathrm{ness}} \equiv \lim_{\epsilon\to 0^+} \hat{U}_{I,\epsilon}(0,-\infty) \rho_0 \hat{U}_{I,\epsilon}^{\dagger}(0,-\infty)$ is given by the exact and intuitive form (see Appendix~\ref{SecSI:NESS} and Refs.~\cite{Zubarev1, Zubarev2, ZubarevBook1, Fujii})
\begin{equation}\label{eq:NESS}
\hat{\rho}_{\mathrm{ness}} = Z_{\mathrm{ness}}^{-1}  e^{-\overline{\beta} \left(\hat{\Ham} -\overline{\mu} \hat{N} \right) + \hat{\Sigma}},
\end{equation}
where $Z_{\mathrm{ness}}$ is the NESS partition function, \GGrev{$N \equiv \sum_{a=L,R} N_a$ is the total particle number operator for the leads}, $\overline{\beta} = 1/2\left(\beta_L+\beta_R\right)$, $\overline{\mu} = \left(\beta_L\mu_L + \beta_R\mu_R\right)/\left(\beta_L+\beta_R\right)$, $\delta_\beta = \beta_L - \beta_R$, and $\delta_{\beta\mu} = \beta_L\mu_L - \beta_R\mu_R$. Crucially, Eq.~(\ref{eq:NESS}) includes the \textit{entropy production operator} $\hat{\Sigma}$ which is defined as
\begin{equation}\label{eq:entrop_pro}
   \hat{\Sigma} = \delta_{\mu\beta} \hat{Q}_+ - \delta_\beta \hat{E}_+,
\end{equation}
where 
\begin{equation}
\hat{E} = \frac{1}{2}\left(\hat{\Ham}_L-\hat{\Ham}_R\right),\qquad \hat{Q} =\frac{1}{2}\left(\hat{N}_L-\hat{N}_R\right),
\end{equation}
and for $X = E, Q$, we introduce the operators
\begin{equation}\label{Xp_identity}
    \hat{X}_+ 
    = \lim\limits_{\epsilon\to0^+} \epsilon \int\limits_{-\infty}^0 dt e^{\epsilon t} \hat{X}_H(t)
    =  \lim_{\mathfrak{T}\to\infty}\frac{1}{\mathfrak{T}}\int_{-\mathfrak{T}}^0 dt \hat{X}_H(t),
\end{equation}
with $\hat{X}_H(t) = e^{i \hat{\Ham} t} \hat{X} e^{-i \hat{\Ham}t}$ and $\mathfrak{T} \propto \epsilon^{-1}$. 
The last equality in Eq.~(\ref{Xp_identity}) is a direct consequence of Abel's theorem~\cite{Fujii, ZubarevBook3}.
$\hat{Q}_+$ and $\hat{E}_+$ therefore represent the operators connected to time-averaged  particle and energy differences between the $L$ and $R$ leads.
It is important to emphasize that in this framework, the NESS state Eq.~(\ref{eq:NESS}) refers to the global state of the composite $LCR$ system and not just the reduced state of $C$.

The structure of Eq.~\eqref{eq:NESS} implies that the NESS is a generalized Gibbs ensemble where in addition to the usual conserved quantities $\hat{\Ham}$ and $\hat{N}$, there are the additional conserved quantities $\hat{E}_+$ and $\hat{Q}_+$, whose Lagrange multipliers are the well known thermodynamic affinities $\delta_{\beta}$ and $\delta_{\beta\mu}$ that drive the energy and particle currents respectively~\cite{de2013non}.
Another observation from Eq.~\eqref{eq:NESS} is that the cumulants of the steady-state entropy production, here connected to the current fluctuations, are generated by taking derivatives of the corresponding Massieu potential $\psi_{\mathrm{ness}} \equiv -\ln Z_{\mathrm{ness}}$~\cite{Taniguchi}. This is in direct analogy to the case for equilibrium ensembles where they are connected to the equilibrium fluctuations of the energy. In Appendix~\ref{SubsecSI:EntropyProd} we explicitly show that the expectation value of the entropy production operator $\hat{\Sigma}$ for the NESS, calculated through the Massieu potential, recovers the familiar result for the \textit{average entropy production rate}
\begin{equation}\label{eq:rate}
    \mean{\hat{\sigma}} = \lim_{\mathfrak{T}\to \infty} \frac{1}{\mathfrak{T}} \mean{\hat\Sigma} = \delta_{\beta\mu} \mean{\hat{J}_Q} - \delta_{\beta} \mean{\hat{J}_E},
\end{equation}
where $\mean{\hat{J}_{Q,E}} = \mathrm{Tr}\left[\hat{J}_{Q,E}\,\hat{\rho}_{\mathrm{ness}}\right]$ are the usual asymptotic steady-state particle and energy currents~\cite{LandauerButtiker, Ness2}, with the current operator of $\hat{X}$ given as $\hat{J}_{X}(t) \equiv \frac{d \hat{X}_H(t)}{dt}$ for $X = E,Q$.

\section{Entropy production as a relative entropy}
\label{sec:entropy}

Armed with the NESS statistical operator we are now ready to introduce our first main result. Using the NESS ensemble Eq.~\eqref{eq:NESS} we write the average entropy production as a quantum relative entropy~\cite{vedral2002role}, which plays a central role in non-equilibrium quantum thermodynamics~\cite{spohn1978entropy,spohn1978irreversible,donald1987free,DeffnerLutz,EspositoNJP2010,deffner2011nonequilibrium}.  
In the formalism presented here, the entropy production operator $\hat{\Sigma}$ defined in Eq.~\eqref{eq:entrop_pro} represents a conserved quantity in the NESS, i.e. it commutes with the total Hamiltonian $\hat\Ham$. Consequently the exponential in Eq.~\eqref{eq:NESS} can be factorized leading to the following insightful re-expression
\begin{equation}\label{eq:NessLes}
    \hat{\rho}_{\mathrm{ness}} = \hat{\rho}_{\mathrm{les}} e^{\hat\Sigma} \frac{Z_{\mathrm{les}}}{Z_{\mathrm{ness}}}, \qquad \hat{\rho}_{\mathrm{les}} \equiv Z_{\mathrm{les}}^{-1} e^{-\overline{\beta} \left(\hat{\Ham} -\overline{\mu} \hat{N} \right)},
\end{equation}
where $\hat{\rho}_{\mathrm{les}}$ represents the \textit{local equilibrium state} (LES) of the total system, i.e. the equilibrium condition that would be reached if both leads had an inverse temperature $\bar{\beta}$ and chemical potential $\bar{\mu}$ [see Fig.~\ref{figStates}(b)].  
The above relation clearly expresses the intimate relation between  $\hat{\rho}_{\mathrm{ness}}$ and $\hat{\rho}_{\mathrm{les}}$. 
This is made precise by quantifying their distinguishability through the relative entropy (Kullback-Leibler divergence) \SCrev{ $D(\hat{\rho}_1||\hat{\rho}_2) = \mathrm{Tr}\left[\hat{\rho}_1\ln\\\hat{\rho}_1\right] - \mathrm{Tr}\left[\hat{\rho}_1\ln\hat{\rho}_2\right]$}. 
A straightforward calculation leads to 
\begin{equation}\label{eq:Result1}
\mean{\hat{\Sigma}}  = D\left(\hat{\rho}_{\mathrm{ness}} || \hat{\rho}_{\mathrm{les}}\right)  + \Delta \psi,
\end{equation}
where the second term on the r.h.s. corresponds to the difference in the Massieu potentials of the LES and of the NESS
\begin{equation}\label{eq:DeltaPsi}
\Delta \psi = \psi_{\mathrm{les}} - \psi_{\mathrm{ness}}   \equiv  \ln\left(\frac{Z_{\mathrm{ness}}}{Z_{\mathrm{les}}}\right) = \ln \left(1+\sum_{n=1}^{+\infty} (2n!)^{-1}\mean{\hat\Sigma^{2n}}_{\mathrm{les}}\right).
\end{equation}
We stress that the expectation value of the even powers of the entropy production operator in the last equality are calculated over the local equilibrium state (see Appendix~\ref{SubsecSI:DeltaPsi} for derivation and details). 
One may physically interpret $\langle W_\mathrm{extr} \rangle = \overline{\beta}^{-1} D(\rho_{\mathrm{ness}}||\rho_{\mathrm{les}})$ as the available work that could be extracted by letting the leads equilibrate to $\overline{\beta}$ and $\overline{\mu}$ ~\cite{NatanAndrei}.

Two relevant considerations can be made.
First, all the expectation values $\mean{\hat{\Sigma}^{2n}}_{\mathrm{les}}$ are positive quantities, therefore leading to the conclusion that $\Delta \psi \geq 0$.
Combining this with the fact that, via Klein's inequality, the relative entropy is non-negative allows us to obtain the non-negativity of the steady-state mean entropy production $\mean{\hat\Sigma}$.
Along the same lines, upon introducing the entropy production rate $\mean{\hat\sigma}$ from Eq.~\eqref{eq:rate} into Eq.~\eqref{eq:Result1}, one obtains an analogous reformulation of this quantity in terms of a relative entropy and also prove its positivity.
Second, in the absence of a temperature gradient, the first term in the series, i.e. $\mean{\hat\Sigma^{2}}_{\mathrm{les}}$, is proportional to the Johnson-Nyquist noise~\cite{Fujii}, which therefore becomes the leading term whenever an expansion over the affinity $\beta (\mu_L-\mu_R)$ is performed.

As a final observation it is important to remark here that Eq.~\eqref{eq:NessLes}, and all the subsequent results, hold true also in the case of arbitrary number of baths. As discussed in Appendix~\ref{SubsecSI:DerivNess1} the only difference for a multiple bath setup is that the explicit form of the entropy production operator $\hat{\Sigma}$ consists of many more terms.

\section{Thermodynamics of precision}
\label{sec:tur}
\begin{figure}[htbp!]
\includegraphics[width=0.8\columnwidth]{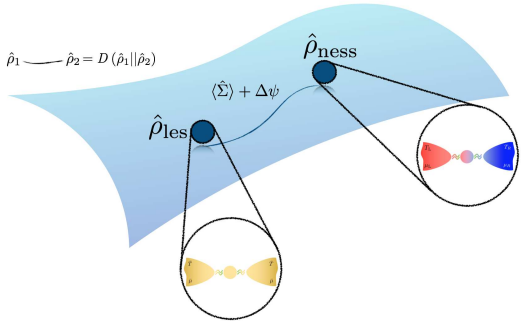}
\caption{(Color online) Manifold of non-equilibrium steady-states and schematic representation of the main results Eqs.~\eqref{eq:Result1}.}
\label{figManifold}
\end{figure}

Using our derived expression for the average entropy production defined by Eq.~\eqref{eq:Result1} we are now in a position to exploit the mathematical properties of the relative entropy to derive our TUR. We start by rewriting the NESS operator in Eq.~\eqref{eq:NESS} in the generic form 
\begin{equation}
\hat{\rho}_{\mathrm{ness}} \equiv \hat{\rho}(\bgreek{\lambda}) = Z_{\mathrm{ness}}^{-1} e^{-\lambda^i \hat{X}_i},
\end{equation}
\SCrev{where $\mathbf{X} = \left(\Ham, -N, E_+, -Q_+\right)^T$ and Einstein's summation notation has been adopted with the vector of parameters $\bgreek{\lambda} = \left( \overline{\beta}, \overline{\beta\mu}, \delta_{\beta}, \delta_{\beta\mu} \right)^T$ representing the set of experimentally controllable conditions defining the manifold of thermodynamically accessible states.} We will henceforth refer to this as the \textit{manifold of steady-states} (SSM), see Fig.~\ref{figManifold}. 

It is immediate to see that equilibrium states of the form $\hat{\rho}_{\rm les}$ will belong to this manifold, as they correspond to vectors $\bgreek{\lambda}$ having the 
affinities $\delta_{\beta}$ and $\delta_{\beta\mu}$ equal to zero. The local curvature of the manifold is then given by the Fisher information, which quantifies the sensitivity of the system to small variations of the control parameter $\bgreek{\lambda}$. Explicitly
\begin{equation}\label{eq:RelEntropyandFisher}
D\left(\hat{\rho} (\bgreek{\lambda}+\delta\bgreek{\lambda}) || \hat{\rho} (\bgreek{\lambda}) \right) = \frac{1}{2} d\bgreek{\lambda}^T \mathbf{I}(\bgreek{\lambda})  d\bgreek{\lambda} + O\left( d\bgreek{\lambda}^3\right),
\end{equation}
where the Fisher information $\mathbf{I}$ is in this multidimensional case a matrix with elements
\begin{equation}
I(\bgreek{\lambda})_{ij}  = \sum_k \rho_k (\bgreek{\lambda}) \left( \frac{\partial \ln \rho_k (\bgreek{\lambda})}{\partial\lambda_i} \frac{ \partial \ln \rho_k (\bgreek{\lambda})}{\partial\lambda_j} \right).
\end{equation}
with $\lbrace\rho_k\rbrace $ denoting the set of populations, i.e. the projections of the density matrix on the energy eigenbasis of $\hat{\Ham}$. The SSM can be shown to be a Riemannian manifold over the set of parameters $\bgreek{\lambda}$, whose metric (called Fubini-Study~\cite{Gibbons1992}) induces a notion of statistical distance between two generic states \SCrev{$\hat{\rho}_1$ and $\hat{\rho}_2$}. In fact looking at the SSM in this geometrical form can be used to generalised the concept of thermodynamic length, introduced by Crooks for the class of equilibrium ensembles~\cite{crooks2007measuring}, to the class of NESS (see also extension to non-unitary dynamics~\cite{scandi2018arxiv}).

In order to shed light on the geometry of the thermodynamics of precision we invoke the Cramer-Rao bound~\cite{kay2013fundamentals} which puts a fundamental lower bound on the precision of estimation of a parameter $\bgreek{\lambda}$, or a function $\mathbf{g}(\bgreek{\lambda})$ of that parameter, labeling a statistical ensemble.
Concretely, the latter reads
\begin{equation}\label{eq:GenCramerRao}
\mathbf{Cov}_{\bgreek{\lambda}}(\mathbf{g})  \geq \mathbf{K}_{\bgreek{\lambda'}}(\mathbf{g})  \mathbf{I}(\bgreek{\lambda})^{-1} \mathbf{K}_{\bgreek{\lambda'}}(\mathbf{g})^T,
\end{equation}
where $\mathbf{K} $ is the Jacobian matrix of transformation with elements  
\begin{equation}
K_{\bgreek{\lambda'}}(\mathbf{g}) _{ij} = \frac{\partial g_i(\bgreek{\lambda)}}{\partial \lambda_j},
\end{equation}
and where $\mathbf{Cov}$ denotes the covariance matrix with elements
\begin{equation}
\mathrm{Cov}_{\bgreek{\lambda}}(g_i g_j) = \mean{ g_i g_j}_{\bgreek{\lambda}} - \mean{ g_i}_{\bgreek{\lambda}} \mean{g_j}_{\bgreek{\lambda}}.
\end{equation}
Equation~\eqref{eq:GenCramerRao} can be interpreted as establishing the positive semi-definiteness of the matrix $\mathbf{Cov} - \mathbf{K} \mathbf{I}^{-1} \mathbf{K}^T$. \SCrev{Given the NESS defined in Eq.~\eqref{eq:NESS} is diagonal in the total energy eigenbasis we can use the classical version of the bound~\cite{paris2009quantum}}. 
\GGrev{We stress that the bound on $\mathbf{Cov}_{\bgreek{\lambda}}(\mathbf{g})$ in Eq.~\eqref{eq:GenCramerRao}, specified by the inverse of the Fisher Information, is valid in general and is in no way restricted to small variation of the control parameter $\bgreek{\lambda}$. Shortly we will perform a series expansion of $\mathbf{Cov} - \mathbf{K} \mathbf{I}^{-1} \mathbf{K}^T$ in powers of $d\bgreek{\lambda}$ which allows the substitution of the Fisher Information with the relative entropy $D\left(\hat{\rho} (\bgreek{\lambda}+\delta\bgreek{\lambda}) || \hat{\rho} (\bgreek{\lambda}) \right) $.}

\GGrev{Let us apply the above general geometrical results Eqs.~\eqref{eq:RelEntropyandFisher} and~\eqref{eq:GenCramerRao} to the situation at hand.}
Motivated by what happens in actual experimental platforms, rather than directly estimating the vector of parameters $\bgreek{\lambda}$, we choose to estimate the average steady-state currents $\mean{\hat{J}_{\alpha}}_{\bgreek{\lambda}} \equiv \mathrm{Tr}\left[ \hat{J}_{\alpha} \hat{\rho}( \bgreek{\lambda}) \right]$. \SCrev{Here we will employ the label $\alpha$ to include all the relevant physical currents, including particle ($\alpha = Q$), energy  ($\alpha = E$), heat  ($\alpha = \Ham$), as well as   the \textit{heat} from the ($\alpha = a = L,R$) reservoir $\hat{J}_{a} = \hat{J}_{E,a} -\mu_a \hat{J}_{Q,a}$ and the work ($\alpha = W$) defined as $\hat{J}_W = \hat{J}_L -\hat{J}_R$.} 
Let us then start from a state $\rho(\bgreek{\lambda}^\star)$ in the SSM corresponding to $\bgreek{\lambda}^\star = \left(\overline{\beta}^\star, \overline{\mu}^\star, 0, 0 \right)^T$, and implement the transformation $\rho(\bgreek{\lambda}^\star) \mapsto \rho(\bgreek{\lambda}^\star + d\bgreek{\lambda}) $, with $d\bgreek{\lambda} = (0, 0, \delta_{\beta}, \delta_{\beta\mu})^T$ being a small increment in the inverse temperature and chemical potential imbalances. These two states represent respectively $\hat{\rho}_{\mathrm{les}}$ and $\hat{\rho}_{\mathrm{ness}}$. 
\GGrev{By performing a series expansion of the Cramer-Rao bound to the leading non-zero order in $d\bgreek{\lambda}$ and exploiting our result Eq.~\eqref{eq:Result1}, we obtain the following inequality (see Appendix~\ref{SubsecSI:QTUR} for details),}
\begin{equation}\label{eq:MAIN2}
\mean{\hat{\sigma}} \geq \mean{\hat{\mathbf{J}}}^T \bgreek{\Delta}^{-1} \mean{\hat{\mathbf{J}}}.
\end{equation}
\SCrev{Here, following the standard procedure~\cite{Barato2015PRL}, we have defined the so-called \emph{normalized} covariance matrix with elements
\begin{equation}\label{eq:covariance}
\Delta_{\hat{J}_\alpha,\hat{J}_\beta} \equiv \lim_{\mathfrak{T}\to \infty} \mathfrak{T} \mathbf{Cov}\left(\hat{J}_\alpha \hat{J}_\beta\right),
\end{equation}
and we have used the entropy production rate defined in Eq.~\eqref{eq:rate} as $\mean{\hat{\sigma}} = \lim_{\mathfrak{T}\to \infty} \mathfrak{T}^{-1} \mean{\hat{\Sigma}}$}. We remind the reader that the \GGrev{adiabatic limit} $\lim_{\mathfrak{T}\to \infty}$ has to be performed at the very end of the calculation of interest~\cite{thirring2013course}, and therefore the ratio in Eq.~\eqref{eq:MAIN2} is well-defined and finite. In particular, from Eq.~\eqref{eq:MAIN2} it immediately follows that
\begin{equation}\label{eq:NewTUR}
\frac{\Delta_{\hat{J}_\alpha}}{\mean{\hat{J}_\alpha}^2}\mean{\hat{\sigma}} \geq 1,
\end{equation}
which represents a TUR of the same form as Eq.~\eqref{classicalTUR} but with a constant which is \GGrev{two} times looser, dictated by the geometry of \SCrev{quantum NESS. Moreover our result in Eq.~\eqref{eq:MAIN2} generalizes the classical TUR since it involves in the full covariance matrix $\bgreek{\Delta}$.}

\section{Implications for meso- and nanoscopic heat engines}
\label{sec:example}

We will now discuss the consequences that our new bound Eq.~\eqref{eq:MAIN2} on precision has on autonomous quantum steady-state machines operating at small biases in temperature and chemical potentials. Let us assume, with reference to the color scheme used in Fig.~\ref{figStates} (a) and without loss of generality, that $T_L > T_R$ and that the thermal gradient is exploited to drive the current against the chemical potential difference $\mu_L < \mu_R$ (Seebeck effect)~\cite{LandauerIBM, LandauerButtiker, Sivan1986PRB, Meir1992PRL}.

In their seminal work, Pietzonka and Seifert (PS) showed that, for steady-state engines described by classical Markovian stochastic processes, the application of TUR to the work current, i.e. the power $\mean{\hat{P}} \equiv \mean{\hat{J}_W} = \mean{\hat{J}_L} -\mean{\hat{J}_R}$ leads straightforwardly to the upper bound Eq.~\eqref{eq:BPS}, with $\eta = \mean{\hat{P}} / \mean{\hat{J}_{L}}$ being the efficiency of the engine and $\eta_C$ being the Carnot efficiency corresponding to $T_L = T_H$ and $T_R = T_C$. A crucial identity that is used to derive Eq.~\eqref{eq:BPS} is the expression of the entropy production rate in terms of the output power and of the efficiency, i.e.
\begin{equation}
    \mean{\hat{\sigma}} = \frac{\mean{\hat{J}_L}}{T_L} - \frac{\mean{\hat{J}_R}}{T_R} = \frac{\mean{\hat{P}}}{T_R} \left(\frac{\eta_C}{\eta}-1\right).
\end{equation}
The bound Eq.~\eqref{eq:BPS} is of paramount importance due to its wide applicability to many systems ranging 
from colloidal systems ~\cite{krishnamurthy2016micrometre} to biological clocks~\cite{barato2016cost}. However, it is expected to fail whenever quantum systems, such as nanoscale heat engines, cannot be suitably described by effective Markovian processes. Indeed recent results~\cite{Agarwalla2018PRB} have in fact shown violations of the classical TUR Eq.~\eqref{classicalTUR} and of the bound on the output power Eq.~\eqref{eq:BPS} in paradigmatic toy models, such as resonant single-dots and serial (or side-coupled) double-dot junctions~\cite{Ptaszynski2018PRB}. 

A straightforward application of our new bound Eq.~\eqref{eq:MAIN2}, when restricted to Eq.~\eqref{eq:NewTUR} with $\hat{J}_\alpha = \hat{J}_W$, leads to
\begin{equation}\label{eq:BGG}
    \mean{\hat{P}} \leq \mathcal{B}_{\mathit{GG}} \equiv \frac{\Delta_{\hat{P}}}{T_R}\left(\frac{\eta_C}{\eta}-1\right) = 2 \mathcal{B}_{\mathit{PS}}.
\end{equation}
This result extends the validity of the conclusions obtained in Ref.~\cite{Pietzonka2018PRL} that were summarized in the introduction of this work. What is more remarkable, however, is that Eq.~(\ref{eq:BGG}) indicates that the allowed output power for given engine efficiency and constancy (i.e. power fluctuations) can potentially be \textit{\GGrev{two} times} larger than any counterpart described as a classical Markov stochastic process. It can moreover be easily checked that all the violations observed in the above mentioned toy models analyzed in Refs.~\cite{Agarwalla2018PRB, Ptaszynski2018PRB} are well within our new bound, even in the presence of Coulomb interaction between the quantum dots ~\cite{Ptaszynski2018PRB} (see also Appendix~\ref{Sec:Example}).

On top of this, an additional bound can be derived by exploiting the full covariance matrix $\bgreek{\Delta}_{\hat{J}_\alpha, \hat{J}_\beta}$ of Eq.~\eqref{eq:MAIN2}. If we consider in particular $\hat{J}_\alpha = \hat{J}_W$ and $\hat{J}_\beta = \hat{J}_L$ (the latter being by convention the heat current from the hot reservoir), we have that (see Appendix~\ref{Sec:Example})
\begin{equation}\label{NewBound}
    \mean{\hat{P}} \leq \frac{\eta}{T_R} \frac{\Delta_{\hat{P}}\Delta_{\hat{J}_L} - \Delta^2_{\hat{P}, \hat{J}_L}}{\Delta_{\hat{P}} - 2\eta\Delta_{\hat{P}, \hat{J}_L} + \eta^2 \Delta_{\hat{J}_L}} \left(\eta_c - \eta\right),
\end{equation}
where $\Delta_{\hat{P}, \hat{J}_L}$ is the normalized covariance between $\hat{P}$ and $\hat{J}_L$ (see Eq.~\eqref{eq:covariance}) and $\Delta_{\hat{J}_L}$ the normalized variance of $\hat{J}_L$.
This new upper bound complements the one of Eq.~\eqref{eq:BGG} and shows an unexpected relation between the maximum amount of power output and the 
incoming heat current from the hot (left) reservoir.
Eq.~\eqref{NewBound} implies that when these two quantities becomes \GGrev{highly statistically correlated}, i.e. $\frac{|\Delta_{\hat{P}, \hat{J}_L}|}{\sqrt{\Delta_{\hat{P}}\Delta_{\hat{J}_L}}} \to 1$, the numerator on the r.h.s. of Eq.~\eqref{NewBound} vanishes and therefore the only way to achieve a finite power output is for the efficiency $\eta$ to become equal to $\eta \to \frac{|\Delta_{\hat{P}, \hat{J}_L}|}{\Delta_{\hat{J}_L}} = \sqrt{\frac{\Delta_{\hat{P}}}{\Delta_{\hat{J}_L}}}$ (so that also the denominator goes to zero). This is the case, for example, in the tight coupling regime, where the heat current becomes proportional to the particle current ~\cite{Esposito2009PRL,Ptaszynski2018PRB}. Since, by definition, $\eta = \mean{\hat{P}}/\mean{\hat{J}_L}$, this means that, for \GGrev{highly statistically correlated} systems, $\mean{\hat{P}} \propto \sqrt{\Delta_P}$ and likewise for the heat current from the hot reservoir. Such relation between the mean values and their variances is typical, e.g., of Gaussian distributions and is expected to hold for ergodic systems. 

\section{Conclusions and Discussions}
\label{sec:conclude}

In this paper we have explored the thermodynamics of precision for quantum NESS. We exploited a statistical ensemble description and an expression of the entropy production in terms of relative entropy in order to bound the dissipation from below by the covariance matrix of currents. Our result differs from the standard approaches in the literature for the thermodynamics of precision -- not only is it derived in a fully quantum mechanical way, it also is geometrical in nature, reflecting the underlying fundamental universality of the concept. 
\GGrev{Moreover, this novel approach the merit of exploiting methods and techniques borrowed from several different research areas, such as quantum information theory, many-body scattering theory and statistical mechanics, and therefore can prove of interest for a wide range of physics community, such as e.g. quantum thermodynamics and condensed matter physics.}

Crucially, the derivation of our result in Eq.~\eqref{eq:MAIN2} does not assume any Markov approximation and it is valid at second order in $\delta_\beta$ and $\delta_{\beta\mu}$, thus beyond the linear response regime. Moreover, as it is the case for the classical TUR, it holds true for any current in the steady-state system. However, it also goes further in that it contains information on the covariance between different currents, whereas the usual TUR that instead concern only the variances of currents, i.e. the diagonal elements of the above covariance matrix. Employing our bound in the context of mesoscopic engines allows us to demonstrate that a machine not modelled with a classical Markovian description can be more powerful. We speculate, from our example that this is due to the presence of quantum coherence. A detailed exploration of this speculation is a study which we are currently undertaking. Additionally, in future work we plan to investigate the repercussions of our bound on the precision of quantum clocks~\cite{erker2017autonomous}, as well as its extension to driven setups~\cite{koyuk2018generalization,Holubec2018PRL}.

\section{Acknowledgments}
\label{sec:acknowledgments}
The authors would like to thank Marti Perarnau-Llobet for insightful comments and Mauro Paternostro, Alessandro Silva, Felix Binder, Paul Riechers, Francesco Plastina, Mark Mitchison and Géraldine Haack for fruitful discussions. This work was supported by an SFI-Royal Society University Research Fellowship (J.G.) This project received funding from the European Research Council (ERC) under the European Union's Horizon 2020 research and innovation program (grant agreement No.~758403). SRC gratefully acknowledges support from the UK’s Engineering and Physical Sciences Research Council (EPSRC) under grant No. EP/P025110/1.
GTL acknowledges the financial support from the S\~ao Paulo Funding Agency (FAPESP) under projects 2017/50304-7, 2017/07973-5.

\appendix 

\section{Derivation of the McLennan-Zubarev generalized statistical operator}
\label{SecSI:NESS}

\subsection{Adiabatic switching of the interaction M{\o}ller operators}
\label{SubsecSI:ASO}

In this section we present some technical details concerning the derivation of the  McLennan-Zubarev generalized statistical ensemble [Eq.~(\ref{eq:NESS}) of the main text], which will be necessary for the derivations of our main results.  
We begin by considering the unitary evolution of a system described subject to a Hamiltonian of the form $\hat{\Ham} = \hat{\Ham}_0 + \hat{\Ham}_{\mathrm{int}}$. 
The Schr\"odinger picture density matrix $\hat{\rho}_S(t)$ will then evolve from some initial time $t_0$ according to $\hat{\rho}_S(t) = \hat{U}(t,t_0)\hat{\rho}_S(t_0) \hat{U}^\dagger(t,t_0)$, where $\hat{U}(t,t_0) = e^{-\ii \hat{\Ham}(t-t_0)}$.
We move to the interaction picture with respect to $\hat{\Ham}_0$ by defining $\hat{\rho}_I(t) = \hat{U}_0^\dagger(t,0) \hat{\rho}_S(t) \hat{U}_0(t,0)$, where $\hat{U}_0(t,0) = e^{-\ii \hat{\Ham}_0 t}$. Note that here we have chosen the time $t=0$, and not $t_0$, as the coincidence time between operators and states in the two pictures.  

The time evolution of $\hat{\rho}_I(t)$ will then be given by $\hat{\rho}_I(t) = \hat{U}_I(t,t_0) \hat{\rho}_I(t_0) \hat{U}_I^{\dagger}(t,t_0)$, where 
\begin{equation}\label{eq:Ui}
    \hat{U}_I(t,t_0) 
    = \hat{U}_0^{\dagger}(t,0) \hat{U}(t,t_0) \hat{U}_0(t_0,0).
\end{equation}
The expectation value of an arbitrary observable $\hat{A}$ not explicitly dependent on time is given by $\langle \hat{A}(t) \rangle = \mathrm{Tr} \left[ \hat{A}_I(t) \hat{\rho}_I(t)\right]$,
where $\hat{A}_I(t) = \hat{U}_0^\dagger(t,0) \hat{A} \hat{U}_0(t,0)$.
In the particular case where  $\hat{\rho}_S(t_0)$ commutes with $\hat{\Ham}_0$, then $\hat{\rho}_S(t_0) = \hat{\rho}_I(t_0) = \hat{\rho}_0$ and $\langle \hat{A}(t) \rangle$ simplifies to 
\begin{equation}
\langle \hat{A}(t) \rangle = \mathrm{Tr}\left[ \hat{A}_I(t) \hat{U}_I(t,t_0)\hat{\rho}_0 \hat{U}_I^\dagger(t,t_0) \right].
\end{equation}
Following standard literature, the steady-state expectation value of $\hat{A}$ is then defined as the asymptotic limit $\langle \hat{A} \rangle_{\rm ness} = \lim_{|t-t_0|\to\infty} \langle \hat{A}(t) \rangle$. We will in particular consider the case where $t = 0$ and $t_0 \to -\infty$. On the one hand, it can be proven that this choice corresponds to taking the correct causal constraint rather than the steady-state corresponding to the advanced solution~\cite{ZubarevBook1, Fujii}. On a more intuitive basis however, one can also motivate this choice on physical grounds by arguing that, when dealing with a steady-state system, one wants to calculate thermodynamic quantities \textit{in} the steady-state, i.e. once the latter is established. This choice is then also consistent with choosing the coincidence time for the quantum mechanical pictures for the evolution to coincide at time $t=0$, where the steady-state is assumed to be reached.
One therefore obtains that 
\begin{equation}\label{SupMat_NESS_def}
\hat{\rho}_\mathrm{ness} := \hat{U}_I(0,-\infty) \,\hat{\rho}_0\,\hat{U}_I^{-1}(0,-\infty).
\end{equation}
We will now specialize this to the McLennan-Zubarev NESS operator in Eq.~(\ref{eq:NESS}) of the main text, and show how it can be constructed starting with $\hat{\rho}_0$ as given by Eq.~(\ref{rho0}) of the main text.
To derive the explicit form of $\hat{\rho}_\mathrm{ness}$ used in Eq.~(\ref{eq:NESS}) of the main text, we next introduce the notion of adiabatic switching of the interaction~\cite{FetterWalecka}.

To that end, we distort the original Hamiltonian to read
\begin{equation}\label{eq:Hamepsilon}
    \hat{\Ham}_{\epsilon}(t) \equiv \hat{\Ham}_0 + e^{-\epsilon |t|} g \hat{\Ham}_{\mathrm{int}},
\end{equation}
with $\epsilon$ being a positive infinitesimal number and $g$ is a dimensionless bookkeeping parameter that can be formally set to unity at the end.
This new Hamiltonian  smoothly interpolates between the  free Hamiltonian $\hat{\Ham}_0$ at $|t| \to \infty$ and the total Hamiltonian $\hat{\Ham}$ at $t = 0$. 
The adiabatic limit corresponds to $\epsilon \to 0^+$, which should be taken only in the end of all  calculations. 
One may now directly verify that for $0 \geq t \geq t_0$, the new interaction picture evolution operator  $\hat{U}_{\epsilon,I}(t,t_0)$ satisfies the differential equation \cite{Molinari}
\begin{equation}\label{remarkableidentity}
        \ii\epsilon g \frac{\partial \hat{U}_{\epsilon,I}(t,t_0)}{\partial g} =
         \hat{\Ham}_{\e,I}(t) \hat{U}_{\epsilon,I}(t,t_0) - \hat{U}_{\epsilon,I}(t,t_0)\hat{\Ham}_{\epsilon,I}(t_0), 
\end{equation}
where $\hat{\Ham}_{\epsilon,I} = \hat{U}^{\dagger}_0(t,0) \hat{\Ham}_{\epsilon} \hat{U}_0(t,0)$.
Specializing to the case where $t = 0$ and $t_0 = -\infty$ and using the fact that $\hat{\Ham}_{\epsilon}(0) = \hat{\Ham}$ and $\hat{\Ham}_{\epsilon}(-\infty) = \hat{\Ham}_0$, we get
\[
\ii\epsilon g \frac{\partial \hat{U}_{\epsilon,I}(0,-\infty)}{\partial g} = \hat{\Ham} \hat{U}_{\epsilon,I}(0,-\infty) - \hat{U}_{\epsilon,I}(0,-\infty) \hat{\Ham}_0.
\]
Taking the limit $\epsilon \to 0$, so the l.h.s. vanishes, and introducing the M{\o}ller operator~(\ref{Moller_def}) defined as 
\begin{equation}\label{Moller_def}
\hat{\Omega}_+ = \lim\limits_{\epsilon \to 0^+} \hat{U}_{\epsilon,I}(0,-\infty),
\end{equation}
then finally leads to the so called  \textit{intertwining property} \GGrev{~\cite{reed1979scattering}}.
\begin{equation}
\label{eq:intertwine}
    \hat{\Ham}\hat{\Omega}_+ = \hat{\Omega}_{+} \hat{\Ham}_0.
\end{equation}
\SCrev{This somewhat counter-intuitive result shows if $\ket{\phi}$ is an eigenstate of $\hat{\Ham}_0$ with energy $E$ then $\hat{\Omega}_+\ket{\phi}$ is an eigenstate of $\hat{\Ham}$ with the same energy. Thus, the M{\o}ller operator formally connects eigenstates of the free and full Hamiltonians.}

\SCrev{An important subtlety of this result, well known from scattering theory ~\cite{kleinert2016particles}, is that although $\Omega_+$ is constructed from a unitary in Eq.~(\ref{Moller_def}) it is rendered {\em non-unitary} in general by the adiabatic limit $\epsilon\to 0^+$. Specifically $\Omega_+$ possesses a left-inverse  $\hat{\Omega}_+^{-1}\hat{\Omega}_+ = \hat{\Id}$ but lacks a right-inverse $\hat{\Omega}_+\hat{\Omega}_+^{-1} \neq \hat{\Id}$. The reason for this is the presence of a discrete set $\mathscr{B}$ of {\em bound states} in the spectrum of $\hat{\Ham}$. The M{\o}ller operator maps the complete continuous spectrum of $\hat{\Ham}_0$ spanning the full Hilbert space $\mathscr{H}$ to only part of the spectrum of $\hat{\Ham}$ spanning the subspace $\mathscr{S}$ of (unbounded) scattering states. Consequently, instead of a right-inverse we strictly have $\hat{\Omega}_+\hat{\Omega}_+^{-1} = \hat{\Pi}_{\mathscr{S}}$, with $\hat{\Pi}_{\mathscr{S}}$ being the projector onto $\mathscr{S}$~\cite{reed1979scattering, thirring2013course}, meaning that Eq.~\eqref{eq:intertwine} can be written as
\begin{equation}
\label{eq:intertwineGellmanproj}
    \hat{\Pi}_{\mathscr{S}}\hat{\Ham}\hat{\Pi}_{\mathscr{S}} = \hat{\Omega}_{+} \hat{\Ham}_0 \hat{\Omega}_{+}^{-1}.
\end{equation}
For clarity in the following we will neglect bound states and assume the unitarity of $\Omega_+^{-1}$ so that instead
\begin{equation}
\label{eq:intertwineGellman}
    \hat{\Ham} = \hat{\Omega}_{+} \hat{\Ham}_0 \hat{\Omega}_{+}^{-1}.
\end{equation}
We will see in the next section that this assumption is tantamount to including bound states in the construction of a generalized Gibbs ensemble for the NESS. However, since bound states by definition do not contribute to currents this is not expected to influence any of our analysis of the NESS.}



\subsection{Derivation of Eq.~\eqref{eq:NESS} of the main text}
\label{SubsecSI:DerivNess1}

We next use Eq.~(\ref{eq:intertwineGellman}) to derive the McLennan-Zubarev ensemble given in Eq.~\eqref{eq:NESS} of the main text.
For any observable $\hat{X}$, we denote $\hat{X}_+ = \hat{\Omega}_+ \hat{X} \hat{\Omega}_+^{-1}$ as the corresponding M{\o}ller evolved operator. 
It is straightforward to check that 
\begin{equation}\label{eq:MollerComm}
    \left[\hat{X},\hat{Y}\right]=0 \Longrightarrow \left[\hat{X}_+,\hat{Y}_+\right] = 0.
\end{equation} 
This property is quite convenient, as the initial state in Eq.~(\ref{SupMat_NESS_def}) is made up of operators that all commute among each other, \GGrev{i.e. $\left[\hat{\Ham}_a, \hat{N}_b\right]=0,\;\; \forall a,b = L,R$}.
Exploiting this, as well as the form of the initial states of $L$ and $R$ in Eq.~(\ref{rho0}) of the main text, we rewrite $\hat{\rho}_\mathrm{ness} = \hat{\Omega}_+\hat{\rho}_0\hat{\Omega}_+^{-1}$ from Eq.~(\ref{SupMat_NESS_def}) as
\begin{equation}\label{rho_NESS_derivation_intermediate_1}
\hat{\rho}_\mathrm{ness} = \prod_{a=L,R}\frac{1}{Z_a} \exp\bigg\{ - \beta_a (\hat{\Ham}_{a,+} - \mu_a \hat{N}_{a,+}) \bigg\} \rho_{C,+}
\end{equation}
We now rearrange the different terms as follows. 
First, we define 
$\bar{\beta} = (\beta_L + \beta_R)/2$,
$\bar{\mu} = (\beta_L \mu_L + \beta_R \mu_R)/(\beta_L + \beta_R)$,
$\delta_\beta = \beta_L - \beta_R$ and $\delta_{\beta\mu} = \beta_L \mu_L - \beta_R \mu_R$. 
Moreover, we define 
\begin{equation}
    \hat{E}_+ \equiv \frac{1}{2}\left(\hat{\Ham}_{L,+}-\hat{\Ham}_{R,+}\right),\qquad \hat{Q}_+ \equiv \frac{1}{2}\left(\hat{N}_{L,+}-\hat{N}_{R,+}\right),
\end{equation}
which are related to the asymptotic values of the energy and particle imbalances, i.e. $\hat{E}_+ = \hat{\Omega}_+ \hat{E} \hat{\Omega}_+^{-1}$ and $\hat{Q}_+ = \hat{\Omega}_+ \hat{Q} \hat{\Omega}_+^{-1}$ with $E \equiv 1/2 (\hat{\Ham}_L - \hat{\Ham}_R)$ and $Q \equiv 1/2 (\hat{N}_L - \hat{N}_R)$. 
We may then write, for instance, 
\begin{align}\label{regroup}
    \beta_L \hat{\Ham}_{L,+} + \beta_R \hat{\Ham}_{R,+} &= \overline{\beta}\left(\hat{\Ham}_{L,+} + \hat{\Ham}_{R,+}\right) + \delta_{\beta}  \hat{E}_+ \notag\\
&= \overline{\beta} \hat{\Ham}  - \overline{\beta} \hat{\Ham}_{C,+} +  \delta_{\beta}  \hat{E}_+ ,
\end{align}
where we used the fact that $\hat{\Ham}_{L,+} + \hat{\Ham}_{R,+} + \hat{\Ham}_{C,+} = \hat{\Ham}$ is the full Hamiltonian. 
A similar result holds for the particle number operators. 
With this rearrangement, Eq.~(\ref{rho_NESS_derivation_intermediate_1}) may now be written as 
\begin{multline}\label{rho_NESS_derivation_intermediate_2}
\hat{\rho}_\mathrm{ness} = \frac{1}{Z_L Z_R}  e^{-\overline{\beta} (\hat{\Ham} - \overline{\mu} \hat{N}) - \delta_\beta \hat{E}_+ + \delta_{\beta\mu} \hat{Q}_+}\\ \times e^{\overline{\beta}(\hat{H}_{C,+} - \mu \hat{N}_{C,+})} \; \hat{\rho}_{C,+} .
\end{multline}

As shown in Refs.~\cite{Fujii,Ness2} the state $\hat\rho_\mathrm{ness}$ is independent of the initial state $\hat{\rho}_C$ of the central system. \GGrev{To prove this is not a trivial task and there are many possible equivalent ways; here we will start by showing that the above mentioned Gell-Mann Low relation
\begin{multline}\label{eq:GLth}
  \hat{U}_{\epsilon,I}(0,-\infty) \hat{\Ham}_0 \hat{U}^{\dagger}_{\epsilon,I}(0,-\infty) = \hat{\Ham} \\
  - \ii\epsilon g \frac{\partial \hat{U}_{\epsilon,I}(0,-\infty)}{\partial g}  \hat{U}^{\dagger}_{\epsilon,I}(0,-\infty) 
\end{multline}
does not depend on the particular partition of the total Hamiltonian $\hat{\mathcal{H}}$ into $\hat{\mathcal{H}}_0$ and $\hat{\mathcal{H}}_{\mathrm{int}}$. With reference to the notation introduced in Section~\ref{sec:ness}, let us consider two scenarios: in the first one $\hat{\Ham}_0 = \hat{\Ham}_L + \hat{\Ham}_R + \hat{\Ham}_C$ and $\hat{\Ham}_{\mathrm{int}} = \hat{V}_{\mathrm{LC}} + \hat{V}_{\mathrm{RC}} $; in the second scenario, which for clarity we will denote with the ``dash'' symbol, $\hat{\Ham}'_0 = \hat{\Ham}_L + \hat{\Ham}_R $ and $\hat{\Ham}'_{\mathrm{int}} =  \hat{\Ham}_C + \hat{V}_{\mathrm{LC}} + \hat{V}_{\mathrm{RC}} $.
The first term on the right hand side of Eq.~\eqref{eq:GLth} contains the full Hamiltonian $\hat{\Ham}$ and therefore does not depend on the particular partition chosen.
Since $\hat{U}^{\dagger}_{\epsilon}(0,-\infty) = \hat{U}^{\dagger}_{\epsilon,I}(0,-\infty)$ and using the explicit expression for the evolution operator 
\begin{align*}
  \hat{U}^{\dagger}_{\epsilon}(0,-\infty) &= \Id + \sum_{n=1}^{+\infty} \frac{(-\ii g)^n}{n!} \int_{-\infty}^0 dt_1 \ldots dt_n e^{-\epsilon\sum_{j=1}^n |t_j|}\notag\\
  &\qquad \times\overrightarrow{\hat{T}}\left[ \hat{\Ham}_{\epsilon}(t_1)\ldots \hat{\Ham}_{\epsilon}(t_n) \right],
\end{align*}
with $\hat{\Ham}_{\epsilon}(t)$ given by Eq.~\eqref{eq:Hamepsilon}, one can easily compute the last term on the right hand side of Eq.~\eqref{eq:GLth} and obtain that 
\begin{align*}
  &\ii\epsilon g \frac{\partial \hat{U}_{\epsilon}(0,-\infty)}{\partial g}  \hat{U}^{\dagger}_{\epsilon}(0,-\infty) \\
  &\qquad =\epsilon 
  \sum_{n=1}^{+\infty} \frac{(-\ii g)^{n-1}}{n!} \int_{-\infty}^{+\infty} dt_1 \ldots dt_n e^{-\epsilon\sum_{j=1}^n |t_j|}\notag\\
  &\qquad\qquad\qquad\qquad\times\overrightarrow{\hat{T}}\left[ \hat{\Ham}_{\epsilon}(t_1)\ldots \hat{\Ham}_{\epsilon}(t_n) \right].
\end{align*}
It is then straightforward to see that $\hat{\Ham}_{\epsilon}(t_j) = \hat{\Ham}'_{\epsilon}(t_j) + \left(1 - e^{-\epsilon |t_j|}\right) \hat{\Ham}_C$ and therefore, since the $\hat{\Ham}_C$ contribution vanishes in the limit $\epsilon \to 0^+, \, g\to 1$, it does not depend on the partitioning of the total Hamiltonian. In turn this implies that the intertwining property Eq.~\eqref{eq:intertwineGellman} is also independent on the chosen partition. }

\SCrev{For the construction of the NESS we required that the initial state $\hat{\rho}_0$ commutes with the chosen free Hamiltonian, e.g. $\hat{\Ham}_0$ or $\hat{\Ham}'_0$. This is satisfied for $\hat{\Ham}_L$ and $\hat{\Ham}_R$ by the assumption of grand-canonical states for the leads made in Eq.~\eqref{rho0}. The freedom to include or not $\hat{\Ham}_C$ in the definition of the free Hamiltonian implies that $\hat{\rho}_{\mathrm{ness}}$ is independent on the choice of $\hat{\rho}_C$ used within $\hat{\rho}_0$. This is a sensible consequence of the fact that the central conductor is only a finite contribution to an otherwise infinite system. It is therefore convenient to exploit this independence by choosing $\hat{\rho}_C$ such that $\hat{\rho}_{C,+} = e^{-\overline{\beta}(\hat{H}_{C,+} - \mu \hat{N}_{C,+})}$, cancelling out the second line of Eq.~\eqref{rho_NESS_derivation_intermediate_2} and reducing it to the McLennan-Zubarev form given in Eq.~\eqref{eq:NESS}.}


\SCrev{Some additional important observations can be made about the operators comprising the McLennan-Zubarev form. First, $[\hat{E},\hat{\Ham}_0]=0$ as well as $[\hat{Q},\hat{\Ham}_0]=0$, which in light of $\hat{\Ham}_{0,+} = \hat{\Ham}$ from Eq.~\eqref{eq:intertwineGellman} and Eq.~\eqref{eq:MollerComm} means that $[\hat{E}_+,\hat{\Ham}]=0$ and $[\hat{Q}_+,\hat{\Ham}]=0$ also. Second, this immediately implies that the entropy production operator $\hat{\Sigma} = \delta_{\mu\beta} \hat{Q}_+ - \delta_\beta \hat{E}_+$ also commutes with $\hat{\Ham}$ and is therefore a conserved quantity.}

It is moreover worth mentioning that, by exploiting the Dyson's expansion of $\hat{U}_{\epsilon,I}(0,-\infty)$ and Abel's theorem~\cite{ZubarevBook1}, it is possible to express $\hat{E}_+$ and $\hat{Q}_+$ as time-averaged Heisenberg-picture operators as $\left( X \equiv E, Q\right)$
\begin{align}\label{eq:Fujiieq}
\hat{X}_+ &= \hat{X} - \int_{-\infty}^0 dt e^{\epsilon t} \hat{J}_X(t), \\
    &= \lim\limits_{\epsilon\to0^+} \epsilon \int\limits_{-\infty}^0 dt e^{\epsilon t} \hat{X}_H(t)
    =  \lim_{\mathfrak{T}\to\infty}\frac{1}{\mathfrak{T}}\int_{-\mathfrak{T}}^0 dt \hat{X}_H(t),
\end{align}
with $\hat{X}_H(t) = e^{i \hat{\Ham} t} \hat{X} e^{-i \hat{\Ham}t}$ and where we have defined the current operator (in Heisenberg picture) of the $\hat{X}$ as $\hat{J}_{X}(t) \equiv \frac{d}{dt}\hat{X}_H(t)$. We refer the interested reader to Ref.~\cite{Fujii} for the details.

Finally, it is important to point out here that the above construction of the NESS statistical operator can be extended to the case of arbitrary number of baths (and even to the case of a continuum of baths, see e.g. Ref.~\cite{ZubarevBook1}), with each assumed to start in grand canonical Gibbs ensemble at time $t=t_0$ at their own inverse temperature $\beta_j$ and chemical potential $\mu_j$. Remarkably, the resulting operator has exactly the same structure and properties of Eq.~\eqref{eq:NESS}, the only difference being the explicit form of the entropy production operator $\hat{\Sigma}$ which instead consists of many more terms. To quickly realize this, it is sufficient to notice that a regrouping of, e.g., $\sum_{j=1}^N \hat{\mathcal{H}}_{j_+}$ in the same spirit of Eq.~\eqref{regroup}, will lead to
\begin{multline}\label{eq:extra}
  \sum_{j=1}^N \hat{\mathcal{H}}_{j_+} =  \overline{\beta}\hat{\mathcal{H}} + \sum_{j=1}^N \hat{\mathcal{H}}_{j_+}\!\left[\frac{N-1}{N}\beta_j - \frac{1}{N}\sum_{k\neq j}^N \beta_k \right],
\end{multline}
where $\overline{\beta} = N^{-1}\sum_{j=1}^N \beta_j $ and the symbol $\sum_{k\neq j}$ denotes a summation over all indices $k$ except the one equal to $j$.

Crucially by exploiting Eq.~\eqref{eq:MollerComm}, with $\hat{X} = \hat{\mathcal{H}}_0 = \sum_{j=1}^N \hat{\mathcal{H}}_j$ and $\hat{Y} = \hat{\mathcal{H}}_k$, it is apparent that the terms appearing in Eq.~\eqref{eq:extra}, beside the first one which will enter in the definition of $\hat{\Sigma}$, still commute with the total Hamiltonian $\hat{\mathcal{H}}$. This allows the exponential of Eq.~\eqref{eq:NESS} to be disentangled even in the multiple-baths case, and the expression in terms of the LES as in Eq.~\eqref{eq:NessLes} to be obtained. The following results therefore holds as well.


\section{Proofs of the results and additional details of the calculations} 
\label{SecSI:Proofs}

\subsection{Discussion of the connection of the expectation value of the entropy production operator $\mean{{\hat{\Sigma}}}$ with the entropy production rate}
\label{SubsecSI:EntropyProd}

In light of the formalism illustrated in detail in the previous Subsection, the NESS statistical operator $\hat{\rho}_{\mathrm{ness}}$ is reached at time $t=0$ through an adiabatic switching on of the coupling at initial time $t_0 = -\infty$, where the initial state was $\hat{\rho}_0$. 
What we defined to be ``entropy production operator'' $\hat{\Sigma} = \delta_{\beta\mu} \hat{Q}_+ - \delta_{\beta\mu} \hat{E}_+$ appearing at the exponent of the NESS state Eq.~\eqref{eq:NESS} is therefore, by construction, a quantity which expresses the dissipated work necessary to \textit{create} the steady-state starting from the factorized state $\hat{\rho}_0$. It is therefore  far from evident why its average over the NESS should correspond to the usual expression considered in the steady-state, where $\mean{\hat{\Sigma}} = \lim_{\mathfrak{T}\to \infty} \mathfrak{T} \mean{\hat{\sigma}}$ with $\mean{\hat{\sigma}}$ being given by ~\cite{seifert2012stochastic} 
\begin{equation}\label{eq:usualsigma}
    \mean{\hat{\sigma}} = \delta_{\beta\mu} \mean{\hat{J}_Q} - \delta_{\beta}\mean{\hat{J}_E},
\end{equation}
where the affinities $\delta_{\beta}$ and $\delta_{\beta\mu}$ have been defined previously and $\mean{\hat{J}_{Q,E}}$ are the steady-state values of the particle and energy currents, respectively. The latter are in fact known to be constant in time and all the above mean values are assumed to be taken with respect to the NESS state, i.e. $\mean{\cdot} \equiv \mathrm{Tr}\left[~\cdot~ \hat{\rho}_{\mathrm{ness}}\right]$.

\begin{figure}[htbp!]
\includegraphics[width=\columnwidth]{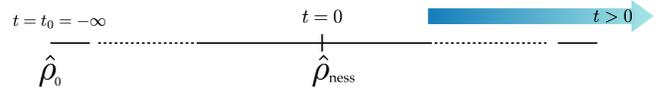}
\caption{(Color online) Schematics of the NESS. The NESS is reached at time $t=0$ through an adiabatic switching on of the interaction at initial time $t_0 = -\infty$, where the state $\hat{\rho}_0$ was factorized. Every expectation value calculated in the steady-state must be then computed at time $t>0$.}
\label{fig2SM}
\end{figure}

Put in another way, can one prove that the mean value of that operator, i.e. the average dissipated work to create the NESS, is actually equivalent to the average entropy production generated in the steady-state (i.e. once the NESS is obtained)?
The answer to this question is {\em affirmative}, and to prove this important fact the first step is to notice that the average currents (particle and energy) in the NESS are time independent.
This was explicitly shown in Ref.~\cite{Fujii} but, in light of its importance, we will repeat here some of the main steps using our notation for convenience.
Let us consider first the expectation value of the particle current at a generic time $t>0$, i.e. once the NESS is established, see Fig.~\ref{fig2SM}
\begin{equation}
    \mean{\hat{J}_Q(t)} = \lim_{\epsilon\to 0^+}\mathrm{Tr}\left[ \hat{U}^{\dagger}_{\epsilon}(t,0) \hat{J}_Q \hat{U}_{\epsilon}(t,0) \hat\rho_{\epsilon} \right],
\end{equation}
where $\hat{U}_{\epsilon}(t,0)$ is the evolution operator corresponding to the Hamiltonian $\hat{\Ham}_{\epsilon}(t)$ and where we have used the notation
\begin{equation}
  \hat{\rho}_{\epsilon} \equiv \hat U_{\epsilon}(0,-\infty)\,\hat\rho_0\,\hat U^{\dagger}_{\epsilon}(0,-\infty).
\end{equation}
Note that the steady-state solution for the statistical operator $\hat{\rho}_{\mathrm{ness}}$ Eq.~\eqref{eq:NESS} is obtained from $\hat{\rho}_{\epsilon}$ by taking the adiabatic limit, i.e. $\hat{\rho}_{\mathrm{ness}} = \lim_{\epsilon\to 0^+} \hat{\rho}_{\epsilon}$.
Finally, in what follows, we will also use the alternative notation $\hat{\rho}_{\mathfrak{T}^{-1}}$ to equivalently denote $\hat{\rho}_{\epsilon}$ after we have switched the limits $\lim_{\epsilon\to 0^+}$ into $\lim_{\mathfrak{T}\to \infty}$ using Abel's theorem (as in Eq.~\eqref{Xp_identity}).

We recall the very important fact that the \GGrev{adiabatic limit} $\lim_{\epsilon\to 0^+}$, which was taken before to get the final closed expression for $\hat{\rho}_{\mathrm{ness}}$, must \textit{always} be performed only at the end of the calculations (i.e. in this case after the average is taken).
By making use of the identity Eq.~\eqref{Xp_identity}, one has that
\begin{equation}\label{eq:Fujii1}
     \mean{\hat{J}_Q(t)} = \lim_{\epsilon\to 0^+}\mathrm{Tr}\left[ \hat{J}_Q \hat{\rho}_{\epsilon} \right] - \int_0^t d\tau\, \mathrm{Tr}\left[ \hat{J}_{\rho_{\epsilon}}(\tau) \hat{J}_Q(t) \right],
\end{equation}
where we have defined the current operator (in Heisenberg picture)
\begin{equation}
    \hat{J}_{\rho_{\epsilon}}(t) \equiv \frac{d}{dt}\hat{\rho}_{\epsilon}(t) = \ii\hat{U}^{\dagger}_{\epsilon}(t,0) \left[\hat{\rho}_{\epsilon},\hat{\Ham}_{\epsilon}(t)\right] \hat{U}_{\epsilon}(t,0).
\end{equation}
It is then possible to show that (see Eqs.~(31) to~(34) of Ref.~\cite{Fujii} for the details), upon defining $\hat{J}'_Q(\tau,t) \equiv \hat{U}_{\epsilon}(\tau,0) \hat{J}_Q(t) \hat{U}^{\dagger}_{\epsilon}(\tau,0)$, one can compute the second term in the above expression
\begin{align}
    &\lim_{\epsilon\to 0^+} \mathrm{Tr}\left[ \hat{J}_{\rho_{\epsilon}}(\tau) \hat{J}_Q(t) \right]\!=\! \lim_{\epsilon\to 0^+} \mathrm{Tr}\left[ \ii \left[\hat{\rho}_{\epsilon}, \hat{\Ham}_{\epsilon}(\tau)\right] \hat{J}'_Q (\tau,t) \right] \!=\! \notag\\
    &\lim_{\epsilon\to 0^+} \epsilon \mathrm{Tr}\left[ \left(\hat{\rho}_{\epsilon}-\hat{\rho}_0\right) \hat{J}'_Q (\tau,t) \right] = 0,
\end{align}
which, substituted back into Eq.~\eqref{eq:Fujii1}, gives
\begin{equation}\label{eq:constantJ}
    \lim_{\epsilon\to 0^+}\mean{\hat{J}_Q(t)} = \lim_{\epsilon\to 0^+}\mean{\hat{J}_Q} \equiv \mean{\hat{J}_Q}.
\end{equation}
Analogous calculations hold for the mean steady-state energy current $\mean{\hat{J}_E}$.
Equipped with these results, we can now consider the entropy production operator $\hat{\Sigma}$ defined in Eq.~\eqref{eq:entrop_pro}. \SCrev{We begin by manipulating $\mean{\hat{\Sigma}}$ as 
\begin{widetext}
\begin{align}
    \mean{\hat{\Sigma}} &= \delta_{\beta\mu} \mean{\hat{Q}_+} - \delta_{\beta}\mean{\hat{E}_+} = -\delta_{\beta\mu}\frac{\partial}{\partial (\delta_{\beta\mu})} \ln Z_{\mathrm{ness}} -\delta_{\beta}\frac{\partial}{\partial (\delta_{\beta})} \ln Z_{\mathrm{ness}}, \notag\\
    &= -\lim_{\epsilon\to 0^+} \left[\delta_{\beta\mu}\frac{\partial}{\partial (\delta_{\beta\mu})} \ln \mathrm{Tr}\left[\hat{\rho}_{\epsilon}\right]  +\delta_{\beta}\frac{\partial}{\partial (\delta_{\beta})} \ln \mathrm{Tr}\left[\hat{\rho}_{\epsilon}\right]\right], \notag\\
    &= -\lim_{\epsilon\to 0^+} \lbrace\delta_{\beta\mu} \mathrm{Tr}\left[ \hat{\rho}_{\epsilon} \left( \hat{Q} - \int_{-\infty}^0 dt\, e^{\epsilon t} \hat{J}_Q(t) \right)\right] +\delta_{\beta}\mathrm{Tr}\left[ \hat{\rho}_{\epsilon} \left( \hat{E} - \int_{-\infty}^0 dt\, e^{\epsilon t} \hat{J}_E(t) \right)\right]\rbrace, \notag
\end{align}
where from the second to the third line Eq.~\eqref{eq:Fujiieq} has been used in place of both $\hat{Q}_+$ and $\hat{E}_+$. Next, we integrate by parts and employ Abel's theorem to get
\begin{align}
     \mean{\hat{\Sigma}} &= -\lim_{\epsilon\to 0^+} \left[\delta_{\beta\mu} \int_{-\infty}^0 dt \, \epsilon e^{\epsilon t} \mean{\hat{Q}(t)} +\delta_{\beta}\int_{-\infty}^0 dt \, \epsilon e^{\epsilon t} \mean{\hat{E}(t)} \right], \notag\\
    &= -\lim_{\mathfrak{T} \to +\infty} \frac{1}{\mathfrak{T}} \left(\delta_{\beta\mu} \int_{-\mathfrak{T}}^0 dt \, \mathrm{Tr}\left[ \hat{Q}(t) \hat{\rho}_{\mathfrak{T}^{-1}}(t) \right] +\delta_{\beta}\int_{-\mathfrak{T}}^0 dt \, \mathrm{Tr}\left[ \hat{E}(t) \hat{\rho}_{\mathfrak{T}^{-1}}(t) \right] \right). \notag
\end{align}
Moving the evolution to Heisenberg picture we then get
\begin{align}
    \mean{\hat{\Sigma}} &= -\lim_{\mathfrak{T} \to +\infty} \frac{1}{\mathfrak{T}} \left(\delta_{\beta\mu} \int_{-\mathfrak{T}}^0 dt \, \mathrm{Tr}\left[ \hat{Q}(t) \hat{U}(t,0) \hat{\rho}_{\mathfrak{T}^{-1}} \hat{U}^{\dagger}(t,0) \right] +\delta_{\beta}\int_{-\mathfrak{T}}^0 dt \, \mathrm{Tr}\left[ \hat{E}(t) \hat{U}(t,0) \hat{\rho}_{\mathfrak{T}^{-1}} \hat{U}^{\dagger}(t,0) \right] \right), \notag\\
    &= -\lim_{\mathfrak{T} \to +\infty} \frac{1}{\mathfrak{T}} \left(\delta_{\beta\mu} \int_{-\mathfrak{T}}^0 dt \, \mathrm{Tr}\left[ \hat{Q}(2t) \hat{\rho}_{\mathfrak{T}^{-1}} \right] +\delta_{\beta}\int_{-\mathfrak{T}}^0 dt \, \mathrm{Tr}\left[ \hat{E}(2t) \hat{\rho}_{\mathfrak{T}^{-1}} \right] \right), \notag
\end{align}
and then by using the definition of the current operator for both $\hat{Q}$ and $\hat{E}$ this becomes
\begin{align}
     \mean{\hat{\Sigma}} &= -\lim_{\mathfrak{T} \to +\infty} \frac{1}{\mathfrak{T}} \left(\delta_{\beta\mu} \int_{-\mathfrak{T}}^0 dt \, \mathrm{Tr}\left[ \left(\int_0^{2t} d\tau \, \hat{J}_Q(\tau) \right)\hat{\rho}_{\mathfrak{T}^{-1}} \right] +\delta_{\beta}\int_{-\mathfrak{T}}^0 dt \, \mathrm{Tr}\left[ \left(\int_0^{2t} d\tau \, \hat{J}_E(\tau) \right) \hat{\rho}_{\mathfrak{T}^{-1}} \right] \right), \notag\\
     &= -\lim_{\mathfrak{T} \to +\infty} \frac{1}{\mathfrak{T}} \int_{-\mathfrak{T}}^0 dt \int_0^{2t} d\tau \left(\delta_{\beta\mu}   \mathrm{Tr}\left[ \hat{J}_Q(\tau)\hat{\rho}_{\mathfrak{T}^{-1}} \right] +\delta_{\beta} \mathrm{Tr}\left[ \hat{J}_E(\tau) \hat{\rho}_{\mathfrak{T}^{-1}} \right] \right). \notag
\end{align}
Finally we exploit Eq.~\eqref{eq:constantJ}, integrate and simplify to arrive at
\begin{align}
    \mean{\hat{\Sigma}} &= -\lim_{\mathfrak{T} \to +\infty} \frac{1}{\mathfrak{T}} \int_{-\mathfrak{T}}^0 dt \int_0^{2t} d\tau \left(\delta_{\beta\mu}   \mathrm{Tr}\left[ \hat{J}_Q \hat{\rho}_{\mathfrak{T}^{-1}} \right] +\delta_{\beta} \mathrm{Tr}\left[ \hat{J}_E \hat{\rho}_{\mathfrak{T}^{-1}} \right] \right), \notag\\
    &= -\lim_{\mathfrak{T} \to +\infty} \frac{1}{\mathfrak{T}} \int_{-\mathfrak{T}}^0 dt \, 2t \left(\delta_{\beta\mu}\mean{\hat{J}_Q} +\delta_{\beta}\mean{\hat{J}_E} \right), \notag\\
    &= \lim_{\mathfrak{T} \to +\infty} \mathfrak{T} \left(\delta_{\beta\mu}\mean{\hat{J}_Q} +\delta_{\beta}\mean{\hat{J}_E} \right) = \lim_{\mathfrak{T} \to +\infty} \mathfrak{T} \mean{\hat{\sigma}}.
\end{align}
\end{widetext}
This concludes the proof, as we can now evaluate the mean entropy production rate in the NESS as
\begin{equation}\label{eq:ratedef}
    \mean{\sigma} = \lim_{\mathfrak{T} \to \infty} \frac{1}{\mathfrak{T}} \mean{\hat{\Sigma}},
\end{equation}
the latter being the NESS average of the entropy production operator appearing in $\hat{\rho}_{\mathrm{ness}}$.}

\subsection{Proof of Eq.~\eqref{eq:DeltaPsi}}
\label{SubsecSI:DeltaPsi}

In order to prove Eq.~\eqref{eq:DeltaPsi} it is useful to keep in mind that, as stated in the main text (and proven e.g. in Refs.~\cite{Bokes1, Heinonen}), the time-averaged entropy production operator $\hat{\Sigma}$ is a conserved quantity and therefore it commutes with the total Hamiltonian $\hat{\Ham}$ (and consequently also with the total number operator $\hat{N}$). This allows to `disentangle' the exponential $\exp\left[ -\overline{\beta}\left(\hat{\Ham} - \overline{\mu} \hat{N}\right) + \hat{\Sigma} \right] = \exp\left[ -\overline{\beta}\left(\hat{\Ham} - \overline{\mu} \hat{N}\right) \right] \exp\left[ \hat{\Sigma} \right]$. We have therefore that
\begin{widetext}
\begin{align}
\Delta\psi &\equiv \ln\left(\frac{Z_{\mathrm{ness}}}{Z_{\mathrm{les}}}\right) = \ln \left(\frac{\mathrm{Tr}\left[ e^{ -\overline{\beta}\left(\hat{\Ham} - \overline{\mu} \hat{N}\right) + \hat{\Sigma} }\right]}{\mathrm{Tr}\left[ e^{ -\overline{\beta}\left(\hat{\Ham} - \overline{\mu} \hat{N}\right) }\right]}\right) =\ln\left(\mathrm{Tr}\left[  \frac{e^{ -\overline{\beta}\left(\hat{\Ham} - \overline{\mu} \hat{N}\right)}}{\mathrm{Tr}\left[ e^{ -\overline{\beta}\left(\hat{\Ham} - \overline{\mu} \hat{N}\right) }\right]}\,\, e^{\hat{\Sigma}}\right] \right)= \ln\left(\mathrm{Tr}\left[  \hat{\rho}_{\mathrm{les}} e^{\hat{\Sigma}}\right]\right).
\end{align}
\end{widetext}
Expanding now the exponential operator $e^{\hat{\Sigma}}$ in Maclaurin series $\sum_{m=0}^{+\infty} \hat{\Sigma}^m /m! $ one is left with
\begin{widetext}
\begin{equation}
\Delta\psi = \ln \left( 1 + \sum_{n=0}^{+\infty} \mathrm{Tr}\left[\hat{\rho}_{\mathrm{les}}\,  \left(\frac{\hat{\Sigma}^{(2n+1)}}{(2n+1)!} \right)\right]+ \mathrm{Tr}\left[\hat{\rho}_{\mathrm{les}}\, \left(\sum_{n=1}^{+\infty} \frac{\hat{\Sigma}^{(2n)}}{(2n)!} \right)\right]\right),
\end{equation} 
\end{widetext}
where the first term comes from the $m=0$ term, i.e. the identity $\Id$, which therefore gives back $\mathrm{Tr}\left[\hat{\rho}_{\mathrm{les}}\right] = 1$, and the remaining terms of the series have been sorted into odd and even powers of $\hat{\Sigma}$. Since the expectation value is taken with respect to the LES statistical operator, the former (i.e. the odd powers of the entropy production operator) vanish and only the even powers survive, thus leading to Eq.~\eqref{eq:DeltaPsi}, i.e.
\begin{equation}
\Delta \psi = \ln \left(1+\sum_{n=1}^{+\infty} (2n!)^{-1}\mean{\hat\Sigma^{2n}}_{\mathrm{les}}\right).
\end{equation}
\GGrev{A number of further considerations can be made at this point, which will be useful in the following, especially in deriving the bound Eq.~\eqref{eq:MAIN2}, as explained in detail in Appendix~\ref{SubsecSI:QTUR}.
First of all, let us start from the identity in Eq.~\eqref{eq:NessLes}
\begin{equation}
\hat{\rho}_{\mathrm{ness}} = \hat{\rho}_{\mathrm{les}} e^{x\hat\Sigma} \frac{Z_{\mathrm{les}}}{Z_{\mathrm{ness}}},
\end{equation}
where we introduced a positive \SCrev{dimensionless} constant $x$ that measures the strength of the affinities $\delta_{\beta}$ and $\delta_{\beta\mu}$, that will prove useful to keep track of the order in the following series expansions; the latter can always be re-absorbed into the definition of $\hat{\Sigma}$.
Notice that, due to Eq.~\eqref{eq:NessLes}, the following identity straightforwardly holds
\begin{equation}\label{ratio1}
\frac{Z_{\mathrm{ness}}}{Z_{\mathrm{les}}} = \langle e^{x\hat\Sigma} \rangle_{\mathrm{les}},
\end{equation}
from which it follows
\begin{equation}\label{relevant1}
x \langle \hat{\Sigma} \rangle_{\mathrm{ness}} = \frac{\hat{\Sigma}\langle e^{x\hat\Sigma} \rangle_{\mathrm{les}}}{\langle e^{x\hat\Sigma} \rangle_{\mathrm{les}}}.
\end{equation}
If we now performs a Taylor expansion of the r.h.s. into powers of $x$, exploiting the fact that $\langle \hat{\Sigma} \rangle_{\mathrm{les}} = 0 $, we finally obtain the following relation
\begin{equation}\label{ExpansionSigma}
x \langle \hat{\Sigma} \rangle_{\mathrm{ness}} = x^2 \langle \hat{\Sigma}^2 \rangle_{\mathrm{les}} + o(x^3), 
\end{equation}
which expresses the fact that average of the square of the entropy production operator calculated on the LES is equal, up to second order in the affinities, to the average entropy production in the NESS.
By finally performing an analogous Taylor expansion on $\Delta \psi$, one also finds that
\begin{equation}\label{DeltaPsiExpansion}
\Delta \psi = \frac{x^2}{2} \langle \hat{\Sigma}^2 \rangle_{\mathrm{les}}  + o(x^3) = \frac{x}{2}\langle \hat{\Sigma} \rangle_{\mathrm{ness}} + o(x^3),
\end{equation}
where Eq.~\eqref{ExpansionSigma} has been exploited in the last step.
}

\subsection{Proof of our bound on thermodynamic precision}
\label{SubsecSI:QTUR}

In this Subsection we will provide the explicit derivation of Eq.~\eqref{eq:MAIN2}. As explained in the main text, the starting point is to perform the following transformation on the manifold of steady-states (SSM) 
\begin{equation}
    \hat{\rho}(\bgreek{\lambda}^\star) \mapsto \hat{\rho}(\bgreek{\lambda}') \equiv \hat{\rho}(\bgreek{\lambda}^\star + d\bgreek{\lambda}), 
\end{equation}
where $\bgreek{\lambda}^\star = \left(\overline{\beta}^\star, \overline{\mu}^\star, 0, 0 \right)^T$ and where $d\bgreek{\lambda} = (0, 0, \delta_{\beta}, \delta_{\beta\mu})^T$ represents a small increment in the inverse temperature and chemical potential imbalances. It is immediate to realize that the two states represent $\hat{\rho}_{\mathrm{les}}$ and $\hat{\rho}_{\mathrm{ness}}$, respectively. 
Let us then employ the generalized Cramer-Rao bound to estimate the average steady-state currents $\mean{\hat{J}_{\alpha}}_{\bgreek{\lambda}}$
\begin{equation}\label{eq:AppB31}
\mathbf{Cov}_{\bgreek{\lambda'}}\left(\mathbf{J}\right) -\mathbf{K}_{\bgreek{\lambda'}}(\mathbf{J})  \mathbf{I}(\bgreek{\lambda'})^{-1} \mathbf{K}_{\bgreek{\lambda'}}(\mathbf{J})^T \geq 0,
\end{equation}
where 
\begin{equation}
K_{\bgreek{\lambda'}}(\mathbf{J}) = \frac{d\mean{\mathbf{J}}}{d\bgreek{\lambda}}
\end{equation}
is the Jacobian matrix and where the covariance matrix has elements 
\begin{multline}
\mathbf{Cov}_{\mathbf{J}}(\bgreek{\lambda'}) _{\alpha\beta} \equiv \mathbf{Cov}\left( \hat{J}_\alpha , \hat{J}_\beta\right) \\= \mathrm{Tr}\left[ \hat{J}_\alpha \hat{J}_\beta \rho(\bgreek{\lambda'})\right] - \mathrm{Tr}\left[ \hat{J}_\alpha \hat{\rho}(\bgreek{\lambda'})\right]\mathrm{Tr}\left[ \hat{J}_\beta \hat{\rho}(\bgreek{\lambda'})\right],
\end{multline} 
\SCrev{with the labels $\alpha,\beta$ being any of the current $Q,E,\Ham,L,R,W$ defined in the main text.}
Equation~\eqref{eq:AppB31} expresses the positive semi-definiteness of the matrix $\mathbf{Cov} - \mathbf{K} \mathbf{I}^{-1} \mathbf{K}^T$ and, could also be alternatively re-written as $ \mathbf{I} - \mathbf{K}^T \mathbf{Cov}^{-1} \mathbf{K} \geq 0$ (as they represent the two Shur complements of a common positive semi-definite block matrix, see e.g. Eq~(6.1.3) of Ref.~\cite{zhang2006schur}). Given that $d\mean{\mathbf{J}} = \mathbf{K} d\bgreek{\lambda}$, let us conveniently expressed the above inequality as
\begin{equation}\label{eq:clear}
     d\bgreek{\lambda}^T \mathbf{I} d\bgreek{\lambda} \geq d\mean{\mathbf{J}}^T \mathbf{Cov}^{-1} d\mean{\mathbf{J}}.
\end{equation}
The next step is to notice that
\begin{equation}
d\mean{\hat{J}_\alpha}_{\bgreek{\lambda}'} \equiv \mean{\hat{J}_\alpha}_{\bgreek{\lambda'}} - \mean{\hat{J}_\alpha}_{\bgreek{\lambda}^\star} = \mean{\hat{J}_\alpha}_{\bgreek{\lambda'}},
\end{equation}
as the last term vanishes (the currents are zero on the LES $\hat{\rho}_{\mathrm{les}} = \hat{\rho}(\bgreek{\lambda}^*)$).
Using this result in Eq~\eqref{eq:clear} leads immediately to 
\begin{equation}\label{eq:S18}
    \mean{\mathbf{J}}^T \mathbf{Cov}^{-1} \mean{\mathbf{J}} \leq d\bgreek{\lambda}^T \mathbf{I} d\bgreek{\lambda} = 2 D \left(\hat{\rho}_{\mathrm{ness}} || \hat{\rho}_{\mathrm{les}} \right) 
\end{equation}
where Eq.~\eqref{eq:RelEntropyandFisher} was used in the last step.
Thanks to our result in Eq.~\eqref{eq:Result1}, we can now substitute the relative entropy in the above expression and obtain
\begin{equation}
\left(\mean{\hat{\Sigma}} - \Delta\psi\right) \geq \frac{1}{2} \mean{\mathbf{J}}^T \mathbf{Cov}^{-1} \mean{\mathbf{J}} .
\end{equation}
\GGrev{It is then important to notice that, in the regime of small temperature and chemical potential biases, one has that the Massieu potential difference $\Delta\psi $ reduces to $x \langle \hat{\Sigma} \rangle_{\mathrm{ness}}/2$ following Eq.~\eqref{DeltaPsiExpansion}}; when plugged into the above Equation this immediately leads to the following bound
\begin{equation}
   \mean{\hat{\Sigma}} \geq \mean{\mathbf{J}}^T \mathbf{Cov}^{-1} \mean{\mathbf{J}}.
\end{equation}
Finally, one can express the above bound in terms of the average steady-state entropy production rate, given the relation
\begin{equation}
    \mean{\Sigma} = \lim_{\mathfrak{T} \to \infty} \mathfrak{T} \mean{\hat{\sigma}}.
\end{equation}
Keeping note that the \GGrev{adiabatic limit} $\lim_{\mathfrak{T}\to\infty}$ must be performed only at the very end of calculations, one retraces the exact same steps as done in standard literature of TUR by introducing the normalized covariance matrix 
\begin{equation}
    \bgreek{\Delta}\left(\hat{J}_\alpha \hat{J}_\beta\right) = \lim_{\mathfrak{T}\to \infty} \mathfrak{T} \mathbf{Cov}\left(\hat{J}_\alpha \hat{J}_\beta\right),
\end{equation}
and the time $\mathfrak{T}$ from the expression before taking the \GGrev{adiabatic limit}. It is straightforward to show that this immediately leads to Eq.~\eqref{eq:MAIN2}. 

Finally, we stress that the above result generalizes the TUR as it involves the full covariance matrix, and it follows in particular that the diagonal elements must be positive as well, from which one obtains
\begin{equation}
\frac{\Delta_{\hat{J}_\alpha}}{\mean{\hat{J}_\alpha}^2}\mean{\hat{\sigma}} \geq 1,
\end{equation}
possessing the same structure as the classical TUR but a bound \GGrev{two} times looser than the Markovian classical counterpart.

\subsection{The double serial quantum dots steady-state engine}
\label{Sec:Example}

We will devote this Subsection to briefly show the application of the TUR-derived upper bound on power in a toy model considered in Refs. ~\cite{Agarwalla2018PRB, Ptaszynski2018PRB}. In particular, we will choose the serial double quantum dots junction model since it has been shown to manifest violations of the TUR even at arbitrary small biases $\delta_\beta$ and $\delta_{\beta\mu}$ when a second order expansion is considered. This corresponds to the regime of validity of our new geometrical TUR.

Let us therefore consider a 1D junction system made of two quantum dots, with energies $E_{L,R}$, coupled to each other coherently through a tunnelling amplitude $\Omega$. The two dots are then respectively hybridized with their corresponding lead with a tunnelling amplitude $t_\alpha$ and chemical potential $\mu_\alpha$. The total Hamiltonian is then
\begin{align}
    \hat{\Ham} &= \sum_{a} E_{a} \hat{c}^{\dagger}_a \hat{c}_a + \Omega \left( \hat{c}^{\dagger}_L \hat{c}_R + \hat{c}^{\dagger}_R  \hat{c}_L\right) + \sum_{a, \mathbf{k}} (\epsilon_{\mathbf{k}} -\mu_a) \hat{b}^{\dagger}_{a, \mathbf{k}} \hat{b}_{a, \mathbf{k}} \notag\\
    & + \sum_{a, \mathbf{k}} \left( t_{a} \hat{b}^{\dagger}_{a, \mathbf{k}}  \hat{c}_a + t_{a}^* \hat{b}_{a, \mathbf{k}}\hat{c}^{\dagger}_a \right), \qquad (a = L,R)
\end{align}
where $\lbrace \hat{c}_{a}, \hat{c}^{\dagger}_{a} \rbrace$ are the annihiliation and creation operators for the quantum dots, while $\lbrace \hat{b}_{a, \mathbf{k}}, \hat{b}^{\dagger}_{a, \mathbf{k}} \rbrace$ are those of the fermionic leads for an eigenstate $\mathbf{k}$ with energy $\epsilon_k$. We also assume, without any loss of generality, that $T_L > T_R$.

In the main text we showed how the TUR implies an upper bound to the power, dictated by its fluctuations $\Delta_P$ and by the efficiency $\epsilon$, according to Eq.~\eqref{eq:BGG}. 
The quantities entering this bound are usually calculated using the formalism of non-equilibrium Green’s function~\cite{LandauerIBM, LandauerButtiker, Sivan1986PRB, Meir1992PRL} and concretely are given, in the wideband limit, by the Landauer-B\"{u}ttiker formulas
\begin{align}
    &\mean{\hat{P}} = \frac{(\mu_R-\mu_L)}{2\pi\hbar}\int_{-\infty}^{+\infty} dE\, \tau(E)\,\left( f_L(E) - f_R(E)\right),\notag\\
    & \Delta_P = \frac{(\mu_R-\mu_L)^2}{2\pi\hbar} \int_{-\infty}^{+\infty} dE\, \tau(E)\,\bigg\{ f_L(E) + f_R(E) \notag\\
    &- 2 f_L(E) f_R(E) -\tau(E)\left[f_L(E) - f_R(E)\right]^2\bigg\},
\end{align}
with $f(E) = \left[ e^{\beta(E-\mu)}+1\right]^{-1}$ being the Fermi-Dirac distribution and $\tau(E)$ being the transmission function. We will assume $\tau(E)$ has the form
\begin{equation}
   \tau(E) = \frac{\Gamma_L \Gamma_R \Omega^2}{\left| \left(E - E_L + i\Gamma_L/2\right)\left(E - E_R + i\Gamma_R/2\right) - \Omega^2 \right|^2}.
\end{equation}
where $\Gamma_{a} = 2\pi |t_{a}|^2 d_a$ is the real part of the dot's self-energy quantifying the coupling of lead $a$ to dot $a$, with $d_a$ denoting the density of states of the lead. Finally, the efficiency of the steady-state engine is given by $\epsilon = \mean{\hat{P}} / \mean{\hat{J}_{L}}$, with 
\begin{equation}
   \mean{\hat{J}_{L}} = \frac{1}{2\pi\hbar}\int_{-\infty}^{+\infty} dE \left(E - \mu_L\right) \tau(E)\,\left( f_L(E) - f_R(E)\right),
\end{equation}
denoting the heat current from the left (hot) reservoir, and
\begin{widetext}
\begin{equation}\label{eq:CorrFunc}
    \Delta(\hat{P},\hat{J}_L) = \frac{\mu_R-\mu_L}{2\pi\hbar} \int_{-\infty}^{+\infty} dE\, \left(E - \mu_L\right)\tau(E)\bigg\{ f_L(E) + f_R(E) - 2 f_L(E) f_R(E) -\tau(E)\left[f_L(E) - f_R(E)\right]^2\bigg\}.
\end{equation}
\end{widetext}

\begin{figure}[htbp!]
\begin{center}
\includegraphics[width=\columnwidth]{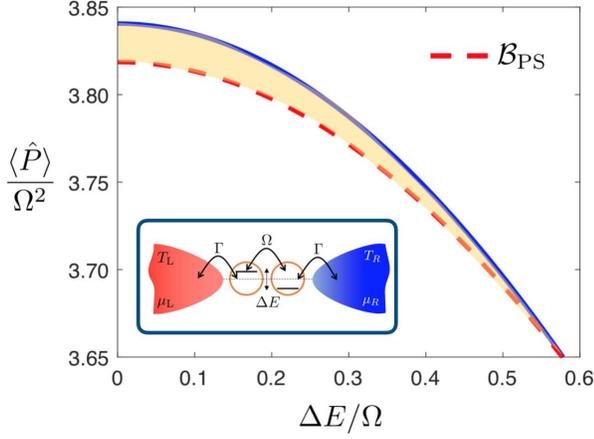}
\caption{(Color online) Plot of the power $\mean{\hat{P}}$ (blue solid curve) against $\mathcal{B}_{PS}$ (red dashed curve) in units of $\Omega^2$ as a function of the two quantum dots' detuning $E_L - E_R = \Delta E$. The other parameters are $ \Gamma_L = \Gamma_R = \Gamma = 0.002 T_R$, $T_L = 10 T_R$, $\mu_L = 1.05 T_R$ $\mu_R = T_R$ and finally $\Omega = \sqrt{15}\Gamma/6$. The inset shows a schematics of the model considered.}
\label{figPowerModel}
\end{center}
\end{figure}

In Fig.~\ref{figPowerModel} the power (blue solid curve) is displayed as a function of the two quantum dots' detuning $E_L - E_R = \Delta E$. In the resonant case $\Delta E = 0$ a Markovian description fails and a violation of the classical TUR arises because of the degeneracy in the system's Hamiltonian. For increasing values of $\Delta E$ the validity of $\mathcal{B}_{PS}$ (red curve) is recovered as sequential hopping becomes dominant. These results are in line with those obtained in Refs.~\cite{Ptaszynski2018PRB, Agarwalla2018PRB} for the same model. Since the violation of the classical TUR in this setup is less than $1\%$ of $\mathcal{B}_{PS}$ it well below our new upper bound $\mathcal{B}_{\mathit{GG}}$ demonstrating that it holds true even in known quantum regimes. Comparisons with the results obtained in Ref.~\cite{Ptaszynski2018PRB} for the violations of $\mathcal{B}_{PS}$ in presence of a Coulomb interaction term in the Hamiltonian $U \hat{c}^{\dagger}_L \hat{c}_L \hat{c}^{\dagger}_R \hat{c}_R $ show that in this case they are also well within the predictions of our bound $\mathcal{B}_{\mathit{GG}}$. 

\subsection{Proof of the new lower bound on power}

Let us start from the re-expression of the entropy production rate in terms of the power and of the efficiency
\begin{equation}
    \mean{\hat{\sigma}} = \frac{\mean{\hat{J}_R}}{T_R} - \frac{\mean{\hat{J}_L}}{T_L} = \frac{\mean{\hat{P}}}{T_R} \left(\frac{\eta_C}{\eta}-1\right).
\end{equation}
Let us choose $\mathbf{\hat{J}} = \left( \hat{J}_W \equiv \hat{P}, \hat{J}_L \right)^T $ (the latter component being by convention the heat current from the hot reservoir). The inverse of the normalized covariance matrix can be calculated explicitly using the following relation, true for any square $n \times n$ matrix $\mathbf{M}$
\begin{equation}
    \mathbf{M}^{-1} = \frac{1}{\mathrm{det}(\mathbf{M})} \mathbf{C}^T, 
\end{equation}
where $\mathbf{C}$ is the square matrix of cofactors of $\mathbf{M}$, i.e. $\mathbf{C}_{ij} = (-1)^{i+j} \mathbf{m}_{ij}$ with $\mathbf{m}_{ij}$ being the minor of $\mathbf{M}$ obtained deleting the $i$th row and $j$th column. The result is given by
\begin{equation}
    \bgreek{\Delta}^{-1} = \frac{1}{\Delta_{\hat{P}}\Delta_{\hat{J}_L} - \Delta^2_{\hat{P}, \hat{J}_L}} \begin{pmatrix} \Delta_{\hat{J}_L} & -\Delta_{\hat{P}, \hat{J}_L} \\ -\Delta_{\hat{P}, \hat{J}_L} & \Delta_{\hat{P}}, \end{pmatrix}
\end{equation}
where we have defined, in conformity of notation with the main text, the normalized correlation function between the power and the heat current from the left (hot) reservoir
\begin{equation}
\bgreek{\Delta}_{\hat{J}_\alpha,\hat{J}_\beta} \equiv \lim_{\mathfrak{T}\to \infty} \mathfrak{T} \mathbf{Cov}\left(\hat{J}_\alpha \hat{J}_\beta\right),
\end{equation}
and 
\begin{equation}
    \Delta_{\hat{J}_\alpha} \equiv \lim_{\mathfrak{T}\to \infty} \mathfrak{T}\left(\langle \hat{J}_\alpha^2\rangle-\langle \hat{J}_\alpha\rangle^2\right)
\end{equation}
the normalized variance of $\hat{J}_{\alpha}$.
The direct application of Eq.~\eqref{eq:MAIN2}, making also use of the definition of the efficiency $\eta \equiv \mean{\hat{P}}/\mean{\hat{J}_H}$, leads straightforwardly to Eq.~\eqref{NewBound}
\begin{equation}
    \mean{\hat{P}} \leq \frac{\eta}{T_R} \frac{\Delta_{\hat{P}}\Delta_{\hat{J}_L} - \Delta^2_{\hat{P}, \hat{J}_L}}{\Delta_{\hat{P}} - 2\eta\Delta_{\hat{P}, \hat{J}_L} + \eta^2 \Delta_{\hat{J}_L}} \left(\eta_c - \eta\right).
\end{equation}

\bibliography{QTUR}

\end{document}